\documentclass[journal]{IEEEtran}

\usepackage{epsfig,amsmath,amssymb,epsf,amsthm,scalefnt,multirow}
\usepackage{xcolor}
\usepackage{float}
\usepackage{cite}
\usepackage{psfrag}
\usepackage{algpseudocode}
\usepackage{algorithm}
\usepackage{soul}
\usepackage{graphicx, subcaption}
\usepackage{makecell}

\newcommand{\argmax}{\mathop{\mathrm{argmax}}}

\newtheorem*{lemma*}{Lemma}

  \def\cC{{\mathcal{C}}}

 \def\cN{{\mathcal{N}}}

\def\argmax{\mathop{\mathrm{argmax}}}
\def\diag{\mathop{\mathrm{diag}}}

\def\b0{{\pmb{0}}} 

\def\ba{{\mathbf{a}}}   \def\bd{{\mathbf{d}}}
 \def\bff{{\mathbf{f}}}  \def\bh{{\mathbf{h}}}
   \def\bl{{\mathbf{l}}}
 \def\bn{{\mathbf{n}}}  \def\bp{{\mathbf{p}}}
   
 \def\bv{{\mathbf{v}}} \def\bw{{\mathbf{w}}} \def\bx{{\mathbf{x}}}

\def\bA{{\mathbf{A}}}   
   \def\bH{{\mathbf{H}}}
\def\bI{{\mathbf{I}}}

\begin{document}

\bstctlcite{IEEEexample:BSTcontrol}

\title{Beam Training for RIS-Aided ISAC Systems}


\author{\IEEEauthorblockN{Jinho Yang, Hyeongtaek Lee, and Junil Choi}
    \thanks{This work was supported in part by the Institute of Information \& Communications Technology Planning \& Evaluation (IITP)-ITRC (Information Technology Research Center) grant funded by the Korea government (MSIT) (IITP-2026-RS-2020-II201787, contribution rate 30\%); in part by Institute of Information \& communications Technology Planning \& Evaluation (IITP) grant funded by the Korea government (MSIT) (No. RS-2024-00395824, Development of Cloud virtualized RAN (vRAN) system supporting upper-midband); and in part by Global - Learning \& Academic research institution for Master’s·PhD students, and Postdocs (G-LAMP) Program of the National Research Foundation of Korea (NRF) grant funded by the Ministry of Education (No. RS-2025-25442252).}
    \thanks{Jinho Yang and Junil Choi are with the School of Electrical Engineering, Korea Advanced Institute of Science and Technology, Daejeon 34141, South Korea (e-mail:\{dwplo3479; junil\}@kaist.ac.kr).}
    \thanks{Hyeongtaek Lee is with the Department of Electronic and Electrical Engineering, Ewha Womans University, Seoul 03760, South Korea (e-mail: htlee@ewha.ac.kr).}
    \thanks{Jinho Yang and Hyeongtaek Lee are co-first authors.}}


\maketitle

\begin{abstract}
    As a key technology for 6G, integrated sensing and communication (ISAC) is receiving considerable attention, and deploying a reconfigurable intelligent surface (RIS) can enhance both communication performance and sensing capability of ISAC by providing additional degrees of freedom. In this paper, we investigate a beam training framework for RIS-aided ISAC systems where beam alignment for a communication user equipment (UE) is conducted while simultaneously detecting a single target through its echo signal.
Using codebooks constructed according to the principles of the 5G standard, we propose a partial search procedure that achieves low training overhead and mathematically show that this strategy is sufficient to identify a suitable codeword combination to serve the UE. By applying the auxiliary beam pair method, the target's angle information from the perspectives of the base station and RIS is obtained. Then, a high-accuracy closed-form localization is proposed based on the angle estimates, and we further extend the proposed technique to multi-target localization scenarios. Numerical results highlight the advantages of the proposed technique in the ISAC context, showing that the training procedure can effectively find a codeword combination and that the target localization technique outperforms the benchmarks.


\end{abstract}

\renewcommand\IEEEkeywordsname{Index Terms}
\begin{IEEEkeywords}
	Integrated sensing and communication (ISAC), reconfigurable intelligent surface (RIS), beam training, target localization.
\end{IEEEkeywords}

\section{Introduction} \label{sec: Introduction}


As represented by 6G, next-generation wireless communication systems aim to realize an even smarter and highly connected wireless environment \cite{6G}. This necessitates both high-accuracy sensing capabilities and high-quality wireless technologies. Consequently, the integrated sensing and communication (ISAC) system is gaining significant attention from both academia and industry as one of the key enabling technologies to meet these demands \cite{ISAC_intro_1, ISAC_intro_2, ISAC_intro_3, ISAC_intro_4}. Although sensing and communication were originally developed independently for different purposes, both technologies have recently adopted high-frequency bands and multiple-input multiple-output (MIMO) systems, leading to similar channel characteristics and signal processing methods \cite{ISAC_intro_3}. This convergence enables the implementation of both sensing and communication functionalities on a single hardware platform, underscoring not only the feasibility but also the necessity of ISAC systems \cite{ISAC_intro_4}.

When an ISAC system tries to improve both sensing and communication performances through additional degrees of freedom, exploiting reconfigurable intelligent surfaces (RISs) will be an effective solution \cite{RIS_ISAC_survey_1, RIS_ISAC_survey_2, RIS_ISAC_survey_3, RIS_ISAC_survey_4, RIS_ISAC_survey_5}. With a deployed RIS, for target sensing, the dual functional radar-communication base station (DFRC-BS) can fully exploit the echo signals both from the BS-target direct link and from the reflection links that consist of BS, target, and RIS. Furthermore, the deployed RIS can enhance the performance of a communication user equipment (UE) through the well-designed BS-RIS-UE link. 

Some prior works studied these advantages of deploying an RIS for the ISAC systems \cite{RIS_ISAC_1, RIS_ISAC_2, RIS_ISAC_4, RIS_ISAC_5, WP-RIS}. In \cite{RIS_ISAC_1}, to maximize the weighted sum of radar signal-to-noise ratios (SNRs) under the communication signal-to-interference-plus-noise ratio constraints, a majorization-minimization based alternating optimization technique was exploited to jointly design the transmit beamforming matrix and RIS reflection coefficients. Instead of the radar SNR, the radar mutual information was maximized under the communication rate constraints in \cite{RIS_ISAC_2} where the proposed technique leveraged the semi-definite relaxation and Riemannian manifold optimization methods. In \cite{RIS_ISAC_4}, the multi-user sum-rate was maximized under the radar sensing constraints based on either the worst-case SNR for target detection or the Cramér–Rao bound for angle estimation. When the unit-modulus constraint is also considered on the transmit waveform at the BS, minimizing the weighted mean squared cross correlation pattern among radar beams under target illumination power and multi-user interference was addressed using the Riemannian manifold optimization in~\cite{RIS_ISAC_5}. More recently, the wireless powered RIS-aided ISAC systems were also investigated by incorporating energy-harvesting constraints, and the beampattern gain was maximized while satisfying communication constraints in \cite{WP-RIS}.

Although many prior works demonstrated notable performance improvements in both sensing and communication, most of the works mainly focused either on theoretical ISAC performance or on designing transmit waveforms at DFRC-BSs with optimized RIS reflection coefficients based on the prior knowledge of target location. However, in order to move beyond theoretical designs and achieve feasible ISAC systems, accurate estimation of the target location becomes necessary in the first place.
Motivated by the above, this paper proposes a framework of RIS-aided ISAC system that enables joint target sensing and beam alignment for a communication UE during the beam training procedure. In realistic communication scenarios, the beam training procedure is essential to identify the optimal combination of codewords at the BS, UE, and RIS that maximizes the communication performance \cite{Beam_training_1, Beam_training_2}. While probing candidate codewords, the DFRC-BS simultaneously receives echo signals reflected from the target, which can be leveraged for target sensing. This enables a seamless integration of sensing capabilities into existing communication systems, thereby motivating the development of a feasible technique that performs target sensing using the predefined codebooks during the beam training procedure. 

A few recent works have investigated practical techniques for realizing RIS-aided ISAC systems during the beam training procedure \cite{RIS_ISAC_beam_training_1, RIS_ISAC_beam_training_2}. In \cite{RIS_ISAC_beam_training_1}, an orthogonal frequency division multiplexing (OFDM) system was considered, and the target angle from the perspective of the BS was estimated using a conventional least-squares approach based on the echo signals received during beam training. The target delay and Doppler frequency were obtained by applying additional wideband signal processing, and the target position was finally estimated. However, the RIS-target link was not utilized to estimate the target angle from the perspective of the RIS, thereby limiting the exploitation of the RIS for target sensing. In \cite{RIS_ISAC_beam_training_2}, the direct path between the BS and the target/UE was assumed to be blocked, and beam training was performed at the RIS under a narrowband scenario. The target angle from the perspective of the RIS was estimated by applying maximum likelihood estimation; however, this approach cannot directly estimate the target position. Motivated by these observations, we develop a framework that fully leverages the RIS to enable target localization even in narrowband systems, e.g., by utilizing one subcarrier in OFDM systems, without relying on additional wideband signal processing. The main contributions of this paper are summarized as follows:
\begin{itemize}
    \item In the case of monostatic sensing using a DFRC-BS architecture, the arrival angles can be estimated from the echo signal reflected by the target after being transmitted from the BS. However, when the distance between the BS and the target is unknown, accurate three-dimensional (3D) localization of the target is generally challenging due to inherent ambiguity. In this context, the RIS can play an important role in overcoming this limitation of ambiguity. When a signal transmitted from the BS is reflected by the RIS and the target, the BS can obtain additional target angle information from the perspective of the RIS. By jointly exploiting the angle information obtained from the perspectives of the BS and RIS, the proposed technique can accurately estimate the 3D position of the target. 
    \item The proper codebooks for the BS and RIS are designed by adopting the codebook design principles of the existing 5G standard where the adjacent elements in a codeword have a constant phase shift. Using these codebooks, the system can select the optimal codeword combination for the communication UE while simultaneously estimating the target’s position, without requiring any additional training resources.
    \item Even when exploiting codebooks designed according to the 5G standard, the resolution of angle estimation will be limited due to the constraints on the number of antennas and available codewords. To tackle this issue, we adopt the auxiliary beam pair (ABP) method, which enables super-resolution angle estimation by exploiting the echo signals obtained from a pair of adjacent beams \cite{ABP, ABP_2}. With the ABP method, we leverage the asymptotic orthogonality of array response vectors to accurately estimate the target angles from the perspective of the BS, even in the presence of additional propagation paths due to the RIS deployment. In addition, we show that the ABP method can also be applied using the RIS reflection coefficients to estimate the target angles from the perspective of the RIS with enhanced resolution. Based on the obtained angle estimates, the proposed localization technique offers a closed-form solution that achieves highly accurate target localization performance. Moreover, we demonstrate that the proposed framework can be extended to multi-target localization.
    \item Numerical results first demonstrate that the optimal codeword combination obtained from the proposed codebook design can achieve a comparable achievable rate performance to that of the baseline case that assumes non-codebook-based beamforming with perfect channel knowledge. Additionally, it is demonstrated that the proposed target localization technique outperforms the other benchmarks in terms of localization accuracy.
\end{itemize}

The rest of this paper is organized as follows. In Section~\ref{sec: System model}, we explain the system model of the RIS-aided monostatic ISAC system. The detailed beam training procedure that we propose is described in Section \ref{sec: Beam training process}. In Section~\ref{sec: Proposed}, the target localization technique based on the ABP method is proposed. Numerical results are shown in Section \ref{sec: Numerical results} to evaluate both the sensing and communication performances of the proposed technique, and the conclusion follows in Section~\ref{sec: Conclusion}.

\textbf{Notations:} Lower and upper boldface letters denote column vectors and matrices. The transpose and conjugate transpose of a matrix $\bA$ are represented by $\bA^\mathrm{T}$ and $\bA^\mathrm{H}$. The diagonalization operation is denoted by $\diag(\cdot)$. Notation $\cC\cN(\boldsymbol{0}_n, \sigma^2\bI_n)$ stands for the complex Gaussian distribution with the mean vector $\boldsymbol{0}_n$ and the covariance matrix $\sigma^2 \bI_n$ where $\boldsymbol{0}_n$ is the $n \times 1$ all-zero vector, and $\bI_n$ denotes the $n \times n$ identity matrix. For a scalar value $a$, $\vert a \vert$ implies the absolute value of $a$. The $\ell_2$-norm of a vector $\ba$ is denoted by $\Vert \ba \Vert_2$. The Kronecker product is denoted by $\otimes$.
\section{System Model} \label{sec: System model}

\begin{figure}[t]
    \centering
    \includegraphics[width=0.8 \columnwidth]{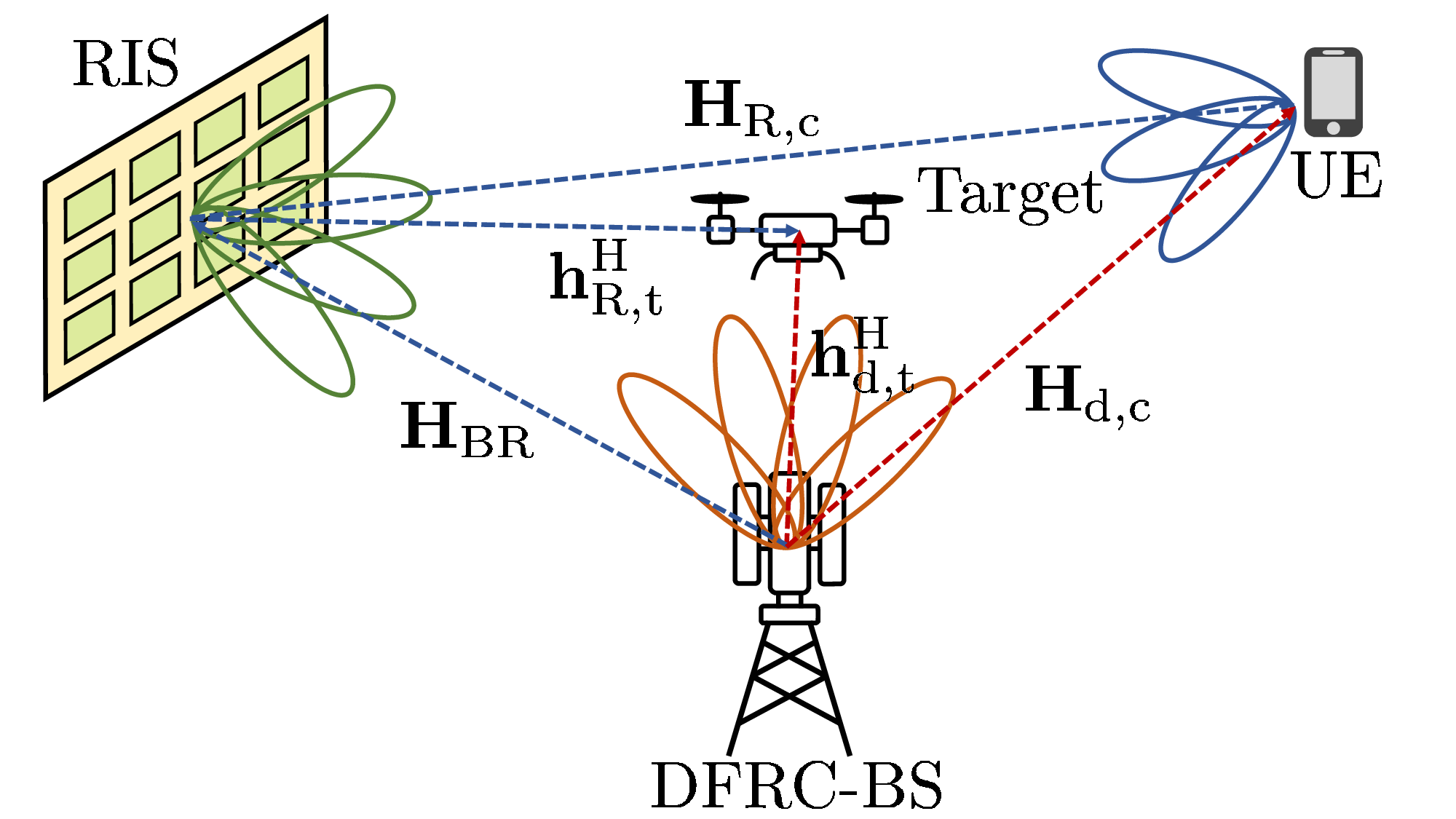}
    \caption{An example of beam training for RIS-aided monostatic ISAC systems.} 
    \label{Fig:System_model}
\end{figure}

We consider a beam training framework\footnote{As in prior works \cite{RIS_ISAC_beam_training_1, RIS_ISAC_beam_training_2, ISAC_beam_training}, we assume that all channels remain fixed during the beam training procedure, which corresponds to scenarios where the UE and target are quasi-static or moderately moving, and the channel coherence time is long enough to complete the beam training.} for an RIS-aided monostatic ISAC system that finds a suitable codeword combination for communication beam alignment by probing candidate codewords, while performing target localization using the echo signals received during the same procedure, as illustrated in Fig.~\ref{Fig:System_model}.
The DFRC-BS equipped with $N$ transmit and receive antennas serves the UE\footnote{The proposed beam training framework can be easily extended to multi-UE scenarios by allowing multiple UEs to simultaneously perform the same procedure where each UE sequentially uses each UE codeword once.} with $L$ antennas and detects a single point-like target by receiving its echo signal.
The RIS, consisting of $M$ passive elements and connected to the BS via a control link, allows the BS to adjust each element to obtain the desired signal reflection.

During the $n$-th time slot, the training signal transmitted by the BS is $s_n \in \mathbb{C}$, which satisfies $\mathbb{E} \left[ \vert s_n \vert^2 \right] = 1$.
The normalized transmit codeword at the BS is $\bff_n \in \mathbb{C}^{N \times 1}$ where $\Vert \bff_n \Vert_2 = 1$.
The reflection coefficient matrix at the RIS is defined as $\boldsymbol{\Phi}_n = \diag(\boldsymbol{\phi}_n)$ where $\boldsymbol{\phi}_n = [\phi_{n, 1}, \cdots, \phi_{n, M}]^\mathrm{T} \in \mathbb{C}^{M \times 1}$ with $\vert \phi_{n, m} \vert = 1$ for $m = 1, \dots, M$.
The normalized receive codeword at the UE is denoted by $\bv_n \in \mathbb{C}^{L \times 1}$ where $\Vert \bv_n \Vert_2 = 1$.
Most previous RIS works assumed that the RIS reflection coefficients are continuously adjustable without any constraints.
However, during the beam training, the reflection coefficients will also be selected from the predefined vectors, which we refer to as the RIS reflection codewords.
Therefore, all codewords including the RIS reflection codeword, i.e., $\bff_n$, $\boldsymbol{\phi}_n$, and $\bv_n$, are selected from the predefined codebooks, which will be elaborated in Section~\ref{sec: Beam training process}.
Under the block fading channel model, the downlink received signal at the UE is
\begin{equation}
    y_{\mathrm{c}, n} = \sqrt{P_\mathrm{T}} \bv_n^\mathrm{H} \left( \bH_{\mathrm{d, c}} + \bH_{\mathrm{R, c}} \boldsymbol{\Phi}_n \bH_{\mathrm{BR}}\right) \bff_n s_n + \bv_n^\mathrm{H} \bn_{\mathrm{c}, n}, \label{eq: received signal at the UE}
\end{equation} 
where $\bH_{\mathrm{d, c}} \in \mathbb{C}^{L \times N}$, $\bH_{\mathrm{R, c}} \in \mathbb{C}^{L \times M}$, and $\bH_{\mathrm{BR}} \in \mathbb{C}^{M \times N}$ represent the communication channels from the BS to the UE, from the RIS to the UE, and from the BS to the RIS, respectively.
The transmit power of the BS is denoted by $P_\mathrm{T}$, and $\bn_{\mathrm{c}, n} \sim \cC\cN \left( \mathbf{0}_L, \sigma_{\mathrm{c}}^2 \bI_L \right)$ is an additive white Gaussian noise (AWGN) vector at the UE.

In the beam training procedure, while the UE receives training signals, the BS simultaneously receives echo signals reflected from the target. For the $n$-th time slot, the normalized receive codeword at the BS is expressed as $\bw_n \in \mathbb{C}^{N \times 1}$ where $\Vert \bw_n \Vert_2 = 1$.
The received echo signal at the BS is given by~\cite{Echo_signal_model_1, Echo_signal_model_2}
\begin{align} \label{echo signal}
    y_{\mathrm{s}, n} & = \sqrt{P_\mathrm{T}} \beta_\mathrm{t} \bw_n^\mathrm{H} \big( \bh_{\mathrm{d, t}} + \bH_{\mathrm{BR}}^\mathrm{H} \boldsymbol{\Phi}_n^\mathrm{H} \bh_{\mathrm{R, t}} \big) \nonumber\\  & \quad \times \big( \bh_{\mathrm{d, t}}^\mathrm{H} + \bh_{\mathrm{R, t}}^\mathrm{H} \boldsymbol{\Phi}_n \bH_{\mathrm{BR}} \big) \bff_n s_n + \bw_n^\mathrm{H} \bn_{\mathrm{s}, n},
\end{align}
where $\bh_{\mathrm{d, t}} \in \mathbb{C}^{N \times 1}$ and $\bh_{\mathrm{R, t}} \in \mathbb{C}^{M \times 1}$ denote the channels from the target to the BS and from the target to the RIS, respectively.
We assume that the normalized radar cross section (RCS) of the target follows the Swerling I model as in \cite{Swerling_model} where $\beta_\mathrm{t} \sim \cC \cN(0, 1)$, and $\bn_{\mathrm{s}, n} \sim \cC\cN \left( \mathbf{0}_N, \sigma_{\mathrm{s}}^2 \bI_N \right)$ is an AWGN vector at the BS.
To align the receive codeword at the BS with the direction of signal transmission, we set $\bw_n = \bff_n$ throughout the beam training procedure, and without loss of generality, we assume the training signal as $s_n = 1$ in the rest of the paper.
Note that self-interference arising in general full-duplex systems can be suppressed using existing self-interference cancellation techniques \cite{self-interference}. In RIS-aided systems, additional self-interference from the BS-RIS-BS path can also be generated, i.e., $\bw_n^\mathrm{H} \bH_{\mathrm{BR}}^\mathrm{H} \boldsymbol{\Phi}_n \bH_{\mathrm{BR}} \bff_n s_n$. We assume that this self-interference component can be mitigated by characterizing the BS-RIS channel using geometric parameters calculated from the known positions of the BS and RIS. This is based on the assumption that the BS-RIS channel includes a dominant line-of-sight (LoS) path because the BS and RIS are fixed and typically installed at high locations. The corresponding self-interference component is then suppressed before processing the echo signals. In addition, when constructing the RIS codebook, RIS reflection codewords that reflect the incident signal toward the BS can be excluded to reduce the residual self-interference. The impact of clutter can be mitigated by estimating clutter characteristics in advance and applying the background subtraction \cite{clutter_subtraction} or filtering \cite{clutter_filtering}.

While our proposed technique does not rely on a specific channel model, we assume that all channels consist of only LoS paths for the sake of clarity.
In Section~\ref{sec: Numerical results}, it will be demonstrated that the proposed technique can still be applied to general scenarios with the dominant LoS path and multiple non-line-of-sight (NLoS) paths.\footnote{Here, we adopt the standard Rician channel model, while the consideration of measurement-based channel characterization and modeling as in \cite{channel_measurement} is an interesting future research direction.}
Considering half-wavelength spacing, the BS antennas and RIS elements are configured as uniform planar array (UPA) structures, and the UE antennas are deployed in a uniform linear array (ULA) structure.
Then, the BS-target and RIS-target channels are
\begin{align}
    &\bh_{\mathrm{d, t}} = \alpha_{\mathrm{d, t}} \ba_\mathrm{B} ( \gamma_{\mathrm{d, t}}, \omega_{\mathrm{d, t}} ), \nonumber \\ & \bh_{\mathrm{R, t}} = \alpha_{\mathrm{R, t}} \ba_\mathrm{R} ( \gamma_{\mathrm{R, t}}, \omega_{\mathrm{R, t}} ),
\end{align}
where $\alpha_{\mathrm{d, t}}$ and $\alpha_{\mathrm{R, t}}$ denote the complex-valued channel gains of the BS-target and RIS-target channels, respectively.\footnote{For simplicity, we include both the path loss and the effect of the number of BS antennas or RIS elements in the complex-valued channel gains.}
The array response vector at the BS is defined as
\begin{equation}\label{eq: array response vector at BS}
    \ba_\mathrm{B} ( \gamma_{\mathrm{d, t}}, \omega_{\mathrm{d, t}} ) = \ba_{\mathrm{B, v}} (\gamma_{\mathrm{d, t}}) \otimes \ba_{\mathrm{B, h}} (\omega_{\mathrm{d, t}}),
\end{equation}
where the vertical and horizontal components are given by
\begin{align}
    &\ba_{\mathrm{B, v}} (\gamma_{\mathrm{d, t}}) = \frac{1}{\sqrt{N_\mathrm{v}}} \left[ 1, e^{j \gamma_{\mathrm{d, t}}}, \cdots, e^{j(N_\mathrm{v}-1) \gamma_{\mathrm{d, t}}} \right]^\mathrm{T}, \nonumber \\ &\ba_{\mathrm{B, h}} (\omega_{\mathrm{d, t}}) = \frac{1}{\sqrt{N_\mathrm{h}}} \left[ 1, e^{j \omega_{\mathrm{d, t}}}, \cdots, e^{j(N_\mathrm{h}-1) \omega_{\mathrm{d, t}}} \right]^\mathrm{T},
\end{align}
where $N_\mathrm{v}$ and $N_\mathrm{h}$ are the number of vertical and horizontal BS antennas that satisfy $N=N_\mathrm{v} \times N_\mathrm{h}$.
The vertical and horizontal arrival spatial frequencies are defined as $\gamma_{\mathrm{d, t}} = \pi \sin(\varphi_{\mathrm{d, t}})$ and $ \omega_{\mathrm{d, t}} = \pi \cos(\varphi_{\mathrm{d, t}}) \sin(\theta_{\mathrm{d, t}})$ where $\varphi_{\mathrm{d, t}}$ and $\theta_{\mathrm{d, t}}$ are the target's vertical and horizontal angles from the perspective of the BS, respectively.
Note that the vertical angle is measured upward from the front direction of the BS, and the horizontal angle is measured counterclockwise from the same reference.
Similarly, the array response vector at the RIS can be written~as
\begin{equation}
    \ba_\mathrm{R} ( \gamma_{\mathrm{R, t}}, \omega_{\mathrm{R, t}} ) = \ba_{\mathrm{R, v}} (\gamma_{\mathrm{R, t}}) \otimes \ba_{\mathrm{R, h}} (\omega_{\mathrm{R, t}}),
\end{equation}
where the vertical and horizontal components are given by
\begin{align}
    &\ba_{\mathrm{R, v}} (\gamma_{\mathrm{R, t}}) = \frac{1}{\sqrt{M_\mathrm{v}}} \left[ 1, e^{j \gamma_{\mathrm{R, t}}}, \cdots, e^{j(M_\mathrm{v}-1) \gamma_{\mathrm{R, t}}} \right]^\mathrm{T}, \nonumber \\& \ba_{\mathrm{R, h}} (\omega_{\mathrm{R, t}}) = \frac{1}{\sqrt{M_\mathrm{h}}} \left[ 1, e^{j \omega_{\mathrm{R, t}}}, \cdots, e^{j(M_\mathrm{h}-1) \omega_{\mathrm{R, t}}} \right]^\mathrm{T},
\end{align}
where $M_\mathrm{v}$ and $M_\mathrm{h}$ are the number of vertical and horizontal RIS elements that satisfy $M=M_\mathrm{v} \times M_\mathrm{h}$.
The vertical and horizontal arrival spatial frequencies are given as $\gamma_{\mathrm{R, t}} = \pi \sin(\varphi_{\mathrm{R, t}})$ and $ \omega_{\mathrm{R, t}} = \pi \cos(\varphi_{\mathrm{R, t}}) \sin(\theta_{\mathrm{R, t}})$ with $\varphi_{\mathrm{R, t}}$ and $\theta_{\mathrm{R, t}}$ representing the target's vertical and horizontal angles from the perspective of the RIS, respectively.
The vertical and horizontal angles are defined in the same way as for the BS, but with the front direction of the RIS as the reference.

Then, the BS-UE, BS-RIS, and RIS-UE communication channels are expressed as
\begin{align}\label{Comm. channels}
    &\bH_{\mathrm{d, c}} = \alpha_{\mathrm{d, c}} \ba_\mathrm{U} ( \nu_\mathrm{d, c} ) \ba_\mathrm{B}^\mathrm{H} ( \gamma_{\mathrm{d, c}}, \omega_{\mathrm{d, c}} ), \nonumber \\ &\bH_{\mathrm{BR}} = \alpha_\mathrm{BR} \ba_\mathrm{R} ( \gamma_\mathrm{RB}, \omega_\mathrm{RB} ) \ba_\mathrm{B}^\mathrm{H} ( \gamma_\mathrm{BR}, \omega_\mathrm{BR} ), \nonumber \\ &\bH_{\mathrm{R, c}} = \alpha_{\mathrm{R, c}} \ba_\mathrm{U} ( \nu_\mathrm{R, c} ) \ba_\mathrm{R}^\mathrm{H} ( \gamma_{\mathrm{R, c}}, \omega_{\mathrm{R, c}} ),
\end{align}
where $\alpha_{\mathrm{d, c}}$, $\alpha_{\mathrm{BR}}$, and $ \alpha_{\mathrm{R, c}}$ are the complex-valued channel gains.
The vertical departure spatial frequencies of each channel are denoted by $\gamma_{\mathrm{d, c}}$, $\gamma_{\mathrm{BR}}$, and $\gamma_{\mathrm{R, c}}$, and the corresponding horizontal departure spatial frequencies are given as $\omega_{\mathrm{d, c}}$, $\omega_{\mathrm{BR}}$, and $\omega_{\mathrm{R, c}}$.
In addition, $\gamma_{\mathrm{RB}}$ and $\omega_{\mathrm{RB}}$ denote the vertical and horizontal arrival spatial frequencies of the BS-RIS channel. 
The array response vector at the UE is defined by 
\begin{equation}
    \ba_\mathrm{U} ( \nu_\mathrm{d, c} ) = \frac{1}{\sqrt{L}} \left[ 1, e^{j \nu_\mathrm{d, c}}, \cdots, e^{j(L-1) \nu_\mathrm{d, c}} \right]^\mathrm{T},
\end{equation}
where $\nu_\mathrm{d, c}$ is the horizontal arrival spatial frequency of the BS-UE channel, and the array response vector $\ba_\mathrm{U} ( \nu_\mathrm{R, c} )$ is defined similarly, using the horizontal arrival spatial frequency of the RIS-UE channel $\nu_\mathrm{R, c}$.
\section{Beam Training Procedure}\label{sec: Beam training process}
Through the beam training procedure in communication systems, an appropriate combination of the transmit codeword at the BS, the RIS reflection codeword, and the receive codeword at the UE will be obtained to maximize communication performance.
During the beam training procedure, the echo signals reflected by the target can be received at the BS, which enables estimation of the target position.
In this section, we first develop the codebook design that is tailored to the simultaneous communication beam training and target localization.
Rather than conducting an exhaustive search over all possible codeword combinations, we propose to adopt a partial search strategy to effectively reduce the training overhead.
At the end of this section, we mathematically demonstrate that the proposed partial search strategy can still achieve sufficient communication performance.

\subsection{Codebook Construction} \label{subsec: Codebook construction}
During the beam training procedure, the proper codeword combination for serving the communication UE is determined, while target localization can also be performed simultaneously.
Therefore, it is necessary to design codebooks that ensure the performance of both functions.
To this end, we consider a codebook structure similar to the discrete Fourier transform (DFT) codebook specified in the 5G standard \cite{DFT_codebook_1}, which features a constant phase shift between adjacent elements of each codeword.
Following this structure, we design codebooks to adopt the ABP method for high-precision angle estimation, enabling accurate target localization, which will be developed in Section \ref{sec: Proposed}.

Since the receive codeword at the UE only affects the communication performance, we adopt the DFT codebook at the UE, which is known to be effective from a communication perspective. Specifically, the UE codebook is expressed as $\mathcal{D}_\mathrm{U} = \left\{ \bv^{(1)}, \cdots, \bv^{(L)} \right\}$ where $\bv^{(i)}$ represent the $i$-th UE codeword. For the BS and RIS codebooks, we design them to follow a structure similar to the DFT codebook, while making it possible to apply the ABP method to estimate the target angles from the perspectives of the BS and RIS.
We first define the BS codebook as $\mathcal{D}_\mathrm{B} = \left\{ \bff^{(1)}, \cdots, \bff^{(N_\mathrm{B})} \right\}$ where $\bff^{(i)}$ denotes the $i$-th codeword for the BS and $N_\mathrm{B}$ is the number of BS codewords. 
To fully utilize the RIS, we first include the transmit codeword at the BS steered toward the RIS, i.e., $\bff^{(1)} \triangleq \ba_\mathrm{B} ( \gamma_\mathrm{BR}, \omega_\mathrm{BR} )$.
Note that $\bff^{(1)}$ can be predefined because the locations of the BS and RIS are fixed.
Based on $\bff^{(1)}$, additional codewords are generated by shifting the vertical and/or horizontal spatial frequencies in integer multiples of $2\delta_\mathrm{B, v}$ and $2\delta_\mathrm{B, h}$, respectively, where $\delta_\mathrm{B, v} = \pi / N_\mathrm{v}$ and $\delta_\mathrm{B, h} = \pi / N_\mathrm{h}$ represent the beam spacings in the vertical and horizontal directions.
Thus, the BS codewords are expressed as 
\begin{equation}
    \bff^{(i)} = \ba_\mathrm{B} ( \gamma_\mathrm{BR} + 2p_\mathrm{B} \delta_\mathrm{B, v}, \omega_\mathrm{BR} + 2q_\mathrm{B} \delta_\mathrm{B, h}),
\end{equation} 
for integers $p_\mathrm{B}$ and $q_\mathrm{B}$.
The range of $p_\mathrm{B}$ is determined to make the vertical spatial frequency $\gamma_\mathrm{BR} + 2p_\mathrm{B} \delta_\mathrm{B, v}$ cover almost all possible range $(-\pi, \pi)$.
The feasible range of $q_\mathrm{B}$ is specified such that the horizontal spatial frequecny $\omega_\mathrm{BR} + 2q_\mathrm{B} \delta_\mathrm{B, h}$ covers the range $ \left(-\pi \sqrt{1 - ( \frac{\gamma_\mathrm{BR} + 2p_\mathrm{B} \delta_\mathrm{B, v}}{\pi})^2}, \pi \sqrt{1 - ( \frac{\gamma_\mathrm{BR} + 2p_\mathrm{B} \delta_\mathrm{B, v}}{\pi})^2} \right)$, which depends on each value of $p_\mathrm{B}$ due to the relationship between the vertical and horizontal spatial frequencies.
This codebook construction results in the number of BS codewords $N_\mathrm{B}$ being smaller than the number of BS antennas $N$.

Similarly, we define the RIS codebook as $\mathcal{D}_\mathrm{R} = \left\{ \boldsymbol{\phi}^{(1)}, \cdots, \boldsymbol{\phi}^{(N_\mathrm{R})} \right\}$ where $\boldsymbol{\phi}^{(i)}$ represents the $i$-th reflection codeword and $N_\mathrm{R}$ denotes the number of RIS reflection codewords.
Let $\delta_\mathrm{R, v} = \pi / M_\mathrm{v}$ and $\delta_\mathrm{R, h} = \pi / M_\mathrm{h}$ be the beam spacings for the vertical and horizontal directions.
Starting from the RIS reflection codeword $\boldsymbol{\phi}^{(1)} \triangleq \sqrt{M} \ba_\mathrm{R} ( \pi - 2 \delta_\mathrm{R, v}, 0 )$, we generate additional reflection codewords by decreasing the vertical spatial frequency in multiples of $2\delta_\mathrm{R, v}$ and/or varying the horizontal spatial frequency in integer multiples of $2\delta_\mathrm{R, h}$.
Each reflection codeword can be represented as 
\begin{equation} \label{reflection_codeword_init}
    \boldsymbol{\phi}^{(i)} = \sqrt{M} \ba_\mathrm{R} ( \pi - 2 p_\mathrm{R} \delta_\mathrm{R, v}, 2 q_\mathrm{R} \delta_\mathrm{R, h}),    
\end{equation}
for a positive integer $p_\mathrm{R}$ and an integer $q_\mathrm{R}$.
The ranges of $p_\mathrm{R}$ and $q_\mathrm{R}$ are determined to cover both possible vertical and horizontal spatial frequency ranges while taking the relationship between vertical and horizontal spatial frequencies into account.
Thus, the number of RIS reflection codewords $N_\mathrm{R}$ is smaller than the number of RIS elements $M$.
To compensate in advance for the arrival spatial frequencies of the BS-RIS link, each reflection codeword is constructed from \eqref{reflection_codeword_init} by shifting the vertical and horizontal spatial frequencies by $-\gamma_\mathrm{RB}$ and $-\omega_\mathrm{RB}$, which is given by
\begin{equation}
    \boldsymbol{\phi}^{(i)} = \sqrt{M} \ba_\mathrm{R} ( \pi - 2 p_\mathrm{R} \delta_\mathrm{R, v} - \gamma_\mathrm{RB}, 2 q_\mathrm{R} \delta_\mathrm{R, h} - \omega_\mathrm{RB}).
\end{equation}

\subsection{Partial Search Procedure} \label{subsec: Partial search}

The simplest way to find the optimal codeword combination from the BS, RIS, and UE is to perform an exhaustive search over all possible combinations.
However, the training overhead of this approach is $N_\mathrm{B} N_\mathrm{R} L$, which can cause excessive delay for initial access.
To address this issue, we propose to adopt a two-stage partial search procedure that significantly reduces the training overhead by evaluating only a subset of all possible codeword combinations.

\begin{figure}[t]
    \centering
    \includegraphics[width=0.85 \columnwidth]{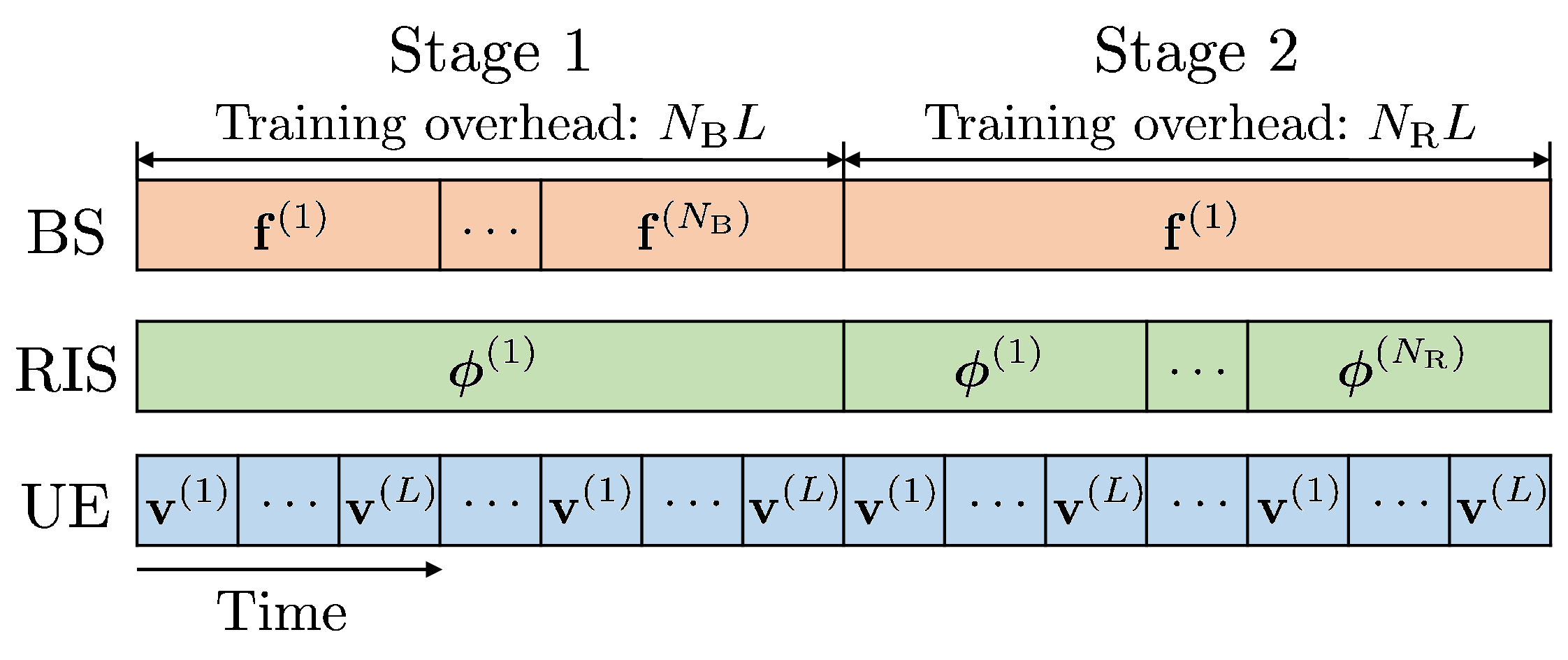}
    \caption{Partial search procedure consisting of two stages.} 
    \label{Fig:Beam_training}
\end{figure}


As will be explained in the next subsection, it is difficult to improve both the BS-UE direct link and BS-RIS-UE reflection link simultaneously in our scenario of interest.
Therefore, in the first training stage, we aim to find the codeword combination best suited for the BS-UE direct link by suppressing the influence of the RIS.
Although the RIS can generally enhance communication performance, in this case, the reflection link may interfere with identifying a suitable codeword combination for the direct link, so the RIS reflection codeword should be set to minimize its impact.
As shown in Fig.~\ref{Fig:Beam_training}, while the RIS reflection codeword is kept fixed, each combination of the BS transmit codeword and UE receive codeword is used once.
During the first stage, we set the RIS reflection codeword as\footnote{While we assume that the RIS is always operating, it is also possible to turn it off during the first stage~\cite{Baseline_MUSIC_Grid, RIS_off}.}
\begin{equation}
     \boldsymbol{\phi}_n = \boldsymbol{\phi}^{(1)} = \sqrt{M} \ba_\mathrm{R} ( \pi - 2 \delta_\mathrm{R, v} - \gamma_\mathrm{RB}, - \omega_\mathrm{RB} ).
\end{equation}
Under the worst-case scenario where the reflected signal is strongest, i.e., when the BS transmit codeword is steered toward the RIS as $\bff_n = \bff^{(1)}$, the reflected signal at the RIS is 
\begin{align}
    \bx_{\mathrm{r},n } &= \diag{(\phi^{(1)})} \bH_{\mathrm{BR}} \bff^{(1)} \nonumber\\ &= \sqrt{M} \alpha_\mathrm{BR} \diag ( \ba_\mathrm{R} ( \pi - 2 \delta_\mathrm{R, v} - \gamma_\mathrm{RB}, - \omega_\mathrm{RB} )) \nonumber \\ &\quad \times \ba_\mathrm{R} ( \gamma_\mathrm{RB}, \omega_\mathrm{RB} ) \ba_\mathrm{B}^\mathrm{H} ( \gamma_\mathrm{BR}, \omega_\mathrm{BR} ) \ba_\mathrm{B} ( \gamma_\mathrm{BR}, \omega_\mathrm{BR} ) \nonumber \\ &= \sqrt{M} \alpha_\mathrm{BR} \ba_\mathrm{R} ( \pi - 2 \delta_\mathrm{R, v}, 0),
\end{align}
which is directed upward from the RIS, toward a direction where the UE and target are unlikely to be located in practical scenarios.
By suppressing the effect of the RIS in this way, the target angles from the perspective of the BS can also be estimated with high precision using the echo signals, with the training overhead of $N_\mathrm{B} L$ in the first stage.

In the second training stage, we search for the codeword combination tailored to the BS-RIS-UE reflection link.
During this stage, the BS continues to transmit the training signal using the fixed transmit codeword $\bff_n = \bff^{(1)}$, while the RIS reflection codeword and the UE receive codeword are varied in different combinations, as illustrated in Fig.~\ref{Fig:Beam_training}.
Since the transmitted signal at the BS is directed toward the RIS, the reflection link becomes more dominant than the direct link; therefore, it is possible to identify the codeword combination most suited for the reflection link. Even if a blockage occurs on the BS-UE direct link, reliable communication can still be maintained through the BS-RIS-UE reflection link using the codeword combination identified during the second stage.
In addition, processing the echo signals received at the BS enables the estimation of the target angles from the perspective of the RIS, and the training overhead of this second stage is~$N_\mathrm{R} L$.

After completing the partial search procedure, the codeword combination is selected to serve the UE.
We first identify the time slot where the downlink received signal at the UE in \eqref{eq: received signal at the UE} is maximized, and then determine the codeword combination used in that time slot.
In addition, the target's position is estimated using the target angles from the perspectives of the BS and RIS obtained after the first and second stages, respectively, which will be elaborated in Section~\ref{sec: Proposed}.
Note that this angle-based localization is efficient in narrowband scenarios because it does not require additional wideband resources or wideband signal processing.
The overall training overhead of the partial search procedure is $(N_\mathrm{B} + N_\mathrm{R}) L$, which is significantly lower than that of the exhaustive search.

\subsection{Performance Analysis for Communication}
In this subsection, we mathematically show that conducting the partial search procedure, rather than the exhaustive search, is sufficient to identify the codeword combination that effectively serves the UE.
By applying the channel models in \eqref{Comm. channels}, the downlink received signal at the UE in \eqref{eq: received signal at the UE} can be expressed~as
\begin{align}\label{eq: received signal at the UE_2}
    y_{\mathrm{c}, n} &= \sqrt{P_\mathrm{T}} \bv_n^\mathrm{H} \left( \bH_{\mathrm{d, c}} + \bH_{\mathrm{R, c}} \boldsymbol{\Phi}_n \bH_{\mathrm{BR}}\right) \bff_n + \bv_n^\mathrm{H} \bn_{\mathrm{c}, n} \nonumber \\ &= \sqrt{P_\mathrm{T}} \bv_n^\mathrm{H} \Big( \alpha_{\mathrm{d, c}} \ba_\mathrm{U} ( \nu_\mathrm{d, c} ) \ba_\mathrm{B}^\mathrm{H} ( \gamma_{\mathrm{d, c}}, \omega_{\mathrm{d, c}} ) \nonumber \\ & \ \ + \alpha_{\mathrm{R, c}} \alpha_\mathrm{BR} \ba_\mathrm{U} ( \nu_\mathrm{R, c} ) \ba_\mathrm{R}^\mathrm{H} ( \gamma_{\mathrm{R, c}}, \omega_{\mathrm{R, c}} ) \boldsymbol{\Phi}_n \nonumber \\ & \ \ \times \ba_\mathrm{R} ( \gamma_\mathrm{RB}, \omega_\mathrm{RB} ) \ba_\mathrm{B}^\mathrm{H} ( \gamma_\mathrm{BR}, \omega_\mathrm{BR} ) \Big) \bff_n + \bv_n^\mathrm{H} \bn_{\mathrm{c}, n}.
\end{align}
Based on \eqref{eq: received signal at the UE_2}, we consider the asymptotic orthogonality of array response vectors, referring to the property that array response vectors steered toward different directions become orthogonal as the number of antennas grows very large.
When the BS transmit codeword is misaligned with the departure spatial frequencies of both the BS-UE and BS-RIS channels, i.e., $\bff_n = \ba_\mathrm{B} ( \gamma_n, \omega_n ) \in \mathcal{D}_\mathrm{B}$ where $(\gamma_n, \omega_n) \neq (\gamma_\mathrm{d, c}, \omega_\mathrm{d, c})$ and $(\gamma_n, \omega_n) \neq ( \gamma_\mathrm{BR}, \omega_\mathrm{BR})$, the following holds~\cite{Orthogonal_1, Orthogonal_2}
\begin{equation}
    \ba_\mathrm{B}^\mathrm{H} ( \gamma_{\mathrm{d, c}}, \omega_{\mathrm{d, c}} ) \bff_n \underset{N \rightarrow \infty}{\approx} 0, \ \ba_\mathrm{B}^\mathrm{H} ( \gamma_\mathrm{BR}, \omega_\mathrm{BR} ) \bff_n \underset{N \rightarrow \infty}{\approx} 0,
\end{equation}
which results in a significant reduction in the downlink received signal at the UE in \eqref{eq: received signal at the UE_2}.
Similarly, when using the UE receive codword that is mismatched to arrival spatial frequencies of both the BS-UE and RIS-UE channels, i.e., $\bv_n = \ba_\mathrm{U} ( \nu_n ) \in \mathcal{D}_\mathrm{U}$ for $\nu_n \neq \nu_{\mathrm{d, c}}$ and $\nu_n \neq \nu_{\mathrm{R, c}}$, the following holds
\begin{equation}
    \bv_n^\mathrm{H} \ba_\mathrm{U} ( \nu_{\mathrm{d, c}} ) \underset{L \rightarrow \infty}{\approx} 0, \ \bv_n^\mathrm{H} \ba_\mathrm{U} ( \nu_{\mathrm{R, c}} ) \underset{L \rightarrow \infty}{\approx} 0, 
\end{equation}
which also degrades the downlink received signal at the UE.

These observations indicate that, due to the asymptotic orthogonality of array response vectors and the single-beam transmission and reception assumption as in our scenario of interest, it is challenging to enhance both the BS-UE direct link and BS-RIS-UE reflection link simultaneously.
For the exhaustive search, which explores all possible codeword combinations, the objective would be to find the best codeword combination that can improve both links at the same time; however, the above limitations imply that this is difficult to achieve.
Therefore, by performing only the partial search procedure that finds the codeword combination tailored to the direct link once and the reflection link once, we can still obtain a sufficiently effective codeword combination.
In addition, the codeword combination obtained from the partial search achieves an achievable rate performance similar to that of the exhaustive search, which will be demonstrated through numerical results in Section~\ref{sec: Numerical results}.


\section{Proposed Target Localization Technique}\label{sec: Proposed}
In this section, we develop a target localization technique by using the echo signals received at the BS during the partial search procedure.
We first introduce the ABP method~\cite{ABP, ABP_2}, which provides high-resolution angle estimates.
To enhance the localization accuracy, we aim to improve the SNR of the echo signals by coherently combining the received echo signals obtained with the same BS transmit codeword and RIS reflection codeword during the partial search.
Then, based on these combined signals, we adopt the ABP method to estimate the target angles from the perspectives of the BS and RIS.
These estimates are obtained through the BS-target direct link and the BS-RIS-target reflection link, respectively.
Using the estimated angles, we propose a closed-form target localization technique and further extend the proposed technique to multi-target localization scenarios.


\subsection{ABP Method}
The angle estimation using the ABP method is performed by comparing the received signals from two spatially adjacent beams.
The two beams are separated by a specific spatial frequency with respect to the central direction, which we call the boresight.
Using the closed-form ratio metric, which can be obtained by comparing the two received signals, this method enables high-precision angle estimation that achieves perfect accuracy in noise-free environments \cite{ABP, ABP_2}.

For easier understanding of the ABP method, in this subsection, we consider a simplified scenario where there is no RIS, and the BS antennas are deployed in the ULA structure.
Then, the received echo signal at the BS in \eqref{echo signal} can be given~as
\begin{align}
    y_{\mathrm{s}, n} & = \sqrt{P_\mathrm{T}} \beta_\mathrm{t} \bw_n^\mathrm{H} \bh_{\mathrm{d, t}} \bh_{\mathrm{d, t}}^\mathrm{H} \bff_n + \bw_n^\mathrm{H} \bn_{\mathrm{s}, n}.
\end{align}
Due to the ULA structure of the BS antennas, the BS-target channel is expressed as
\begin{equation}
    \bh_{\mathrm{d, t}} = \alpha_\mathrm{d, t} \ba_\mathrm{B} ( \nu_\mathrm{d, t} ),
\end{equation}
where the array response vector at the BS is defined by 
\begin{equation}
    \ba_\mathrm{B} ( \nu_\mathrm{d, t} ) = \frac{1}{\sqrt{N}} \left[ 1, e^{j \nu_\mathrm{d, t}}, \cdots, e^{j(N-1) \nu_\mathrm{d, t}} \right]^\mathrm{T},
\end{equation}
with the horizontal arrival spatial frequency $\nu_\mathrm{d, t} = \pi \sin(\psi_\mathrm{d, t})$ where $\psi_\mathrm{d, t}$ represents the target's horizontal angle from the perspective of the BS.

After beam training, we first identify the transmit codeword at the BS that maximizes the received echo signal, which can be denoted as $\bff_n = \ba_\mathrm{B}(\nu_\mathrm{B}^\mathrm{max})$.
This beam (codeword) will be one of the ABP.
Then, based on the identified beam, two spatially adjacent beams, which are separated by a spatial frequency $2 \delta_\mathrm{B}$ where $\delta_\mathrm{B} = \pi/N$ denotes the beam spacing, i.e., $\ba_\mathrm{B}(\nu_\mathrm{B}^\mathrm{max} + 2 \delta_\mathrm{B})$ and $\ba_\mathrm{B}(\nu_\mathrm{B}^\mathrm{max} - 2 \delta_\mathrm{B})$, will be the possible candidates for the other beam to form the ABP.
Among the candidates, the beam with the larger received echo signal will be selected.
For notational simplicity, we define the boresight of the ABP, which is the central direction of the ABP beams, as either $\overline{\nu}_\mathrm{B} = \nu_\mathrm{B}^\mathrm{max} + \delta_\mathrm{B}$ or $\overline{\nu}_\mathrm{B} = \nu_\mathrm{B}^\mathrm{max} - \delta_\mathrm{B}$, depending on the selected beams. 
Then, the ABP beams can be represented as $\bff_n = \ba_\mathrm{B} ( \overline{\nu}_\mathrm{B} - \delta_\mathrm{B} )$ and $\bff_n = \ba_\mathrm{B} ( \overline{\nu}_\mathrm{B} + \delta_\mathrm{B} )$.
The received echo signals when using the ABP beams are written~as
\begin{align}
    y_\mathrm{B}^\Delta &= \sqrt{P_\mathrm{T}} \beta_\mathrm{t} \vert \alpha_\mathrm{d, t} \vert^2 \ba_\mathrm{B}^\mathrm{H} ( \overline{\nu}_\mathrm{B} - \delta_\mathrm{B} ) \ba_\mathrm{B} ( \nu_\mathrm{d, t} ) \nonumber \\& \ \ \times \ba_\mathrm{B}^\mathrm{H} ( \nu_\mathrm{d, t} ) \ba_\mathrm{B} ( \overline{\nu}_\mathrm{B} - \delta_\mathrm{B} ) + \ba_\mathrm{B}^\mathrm{H} ( \overline{\nu}_\mathrm{B} - \delta_\mathrm{B} ) \bn_{\mathrm{s}, n}, \label{received echo a}
    \\ y_\mathrm{B}^\Sigma &= \sqrt{P_\mathrm{T}} \beta_\mathrm{t} \vert \alpha_\mathrm{d, t} \vert^2 \ba_\mathrm{B}^\mathrm{H} ( \overline{\nu}_\mathrm{B} + \delta_\mathrm{B} ) \ba_\mathrm{B} ( \nu_\mathrm{d, t} ) \nonumber \\& \ \ \times \ba_\mathrm{B}^\mathrm{H} ( \nu_\mathrm{d, t} ) \ba_\mathrm{B} ( \overline{\nu}_\mathrm{B} + \delta_\mathrm{B} ) + \ba_\mathrm{B}^\mathrm{H} ( \overline{\nu}_\mathrm{B} + \delta_\mathrm{B} ) \bn_{\mathrm{s}, n}. \label{received echo b}
\end{align}
With the assumption that noise is negligible compared to the desired signal component, the magnitude of the received echo signal in~\eqref{received echo a} can be approximated by
\begin{align}
    \vert y_\mathrm{B}^\Delta \vert & \approx \sqrt{P_\mathrm{T}} \vert \beta_\mathrm{t} \vert \vert \alpha_\mathrm{d, t} \vert^2 \left\vert \ba_\mathrm{B}^\mathrm{H} ( \overline{\nu}_\mathrm{B} - \delta_\mathrm{B} ) \ba_\mathrm{B} ( \nu_\mathrm{d, t} ) \right\vert^2 \nonumber \\ & = \sqrt{P_\mathrm{T}} \vert \beta_\mathrm{t} \vert \vert \alpha_\mathrm{d, t} \vert^2 \left\vert \frac{1}{N} \sum_{k=0}^{N-1} e^{j k (\nu_{\mathrm{d, t}} - \overline{\nu}_\mathrm{B} + \delta_\mathrm{B}) }\right\vert^2 \nonumber \\ &= \sqrt{P_\mathrm{T}} \vert \beta_\mathrm{t} \vert \vert \alpha_\mathrm{d, t} \vert^2 \frac{\cos^2 \left( \frac{N (\nu_{\mathrm{d, t}} - \overline{\nu}_\mathrm{B} )}{2} \right)}{N^2 \sin^2 \left( \frac{\nu_{\mathrm{d, t}} - \overline{\nu}_\mathrm{B} + \delta_\mathrm{B}}{2} \right)}. \label{approx a}
\end{align}
Similarly, we can approximate the magnitude of the received echo signal in~\eqref{received echo b} as
\begin{equation}
    \vert y_\mathrm{B}^\Sigma \vert \approx \sqrt{P_\mathrm{T}} \vert \beta_\mathrm{t} \vert \vert \alpha_\mathrm{d, t} \vert^2 \frac{\cos^2 \left( \frac{N (\nu_{\mathrm{d, t}} - \overline{\nu}_\mathrm{B} )}{2} \right)}{N^2 \sin^2 \left( \frac{\nu_{\mathrm{d, t}} - \overline{\nu}_\mathrm{B} - \delta_\mathrm{B}}{2} \right)}. \label{approx b}
\end{equation} 
Using \eqref{approx a} and \eqref{approx b}, the ratio metric $\xi_\mathrm{B}$ can be defined as
\begin{align}\label{ratio metric}
    \xi_\mathrm{B} &= \frac{\vert y_\mathrm{B}^\Delta \vert - \vert y_\mathrm{B}^\Sigma \vert}{\vert y_\mathrm{B}^\Delta \vert + \vert y_\mathrm{B}^\Sigma \vert} \nonumber = \frac{\sin^2 \left( \frac{\nu_{\mathrm{d, t}} - \overline{\nu}_\mathrm{B} - \delta_\mathrm{B}}{2} \right) - \sin^2 \left( \frac{\nu_{\mathrm{d, t}} - \overline{\nu}_\mathrm{B} + \delta_\mathrm{B}}{2} \right)}{\sin^2 \left( \frac{\nu_{\mathrm{d, t}} - \overline{\nu}_\mathrm{B} - \delta_\mathrm{B}}{2} \right)+ \sin^2 \left( \frac{\nu_{\mathrm{d, t}} - \overline{\nu}_\mathrm{B} + \delta_\mathrm{B}}{2} \right)} \nonumber \\ &= - \frac{\sin(\nu_{\mathrm{d, t}} - \overline{\nu}_\mathrm{B}) \sin(\delta_\mathrm{B})}{1 - \cos(\nu_{\mathrm{d, t}} - \overline{\nu}_\mathrm{B} ) \cos(\delta_\mathrm{B}) }.
\end{align}
As shown in \cite{ABP}, when the target direction lies between the beams in the ABP, i.e., $\vert \nu_{\mathrm{d, t}} - \overline{\nu}_\mathrm{B} \vert < \delta_\mathrm{B}$, the ratio metric $\xi_\mathrm{B}$ is monotonically decreasing and invertible with respect to $\nu_{\mathrm{d, t}} - \overline{\nu}_\mathrm{B}$.
Therefore, by leveraging the inverse function, we can estimate the horizontal arrival spatial frequency as 
\begin{align}
    & \widehat{\nu}_{\mathrm{d, t}} = \overline{\nu}_\mathrm{B} \nonumber \\ & \ \ - \arcsin\left( \frac{\xi_\mathrm{B} \sin (\delta_\mathrm{B}) - \xi_\mathrm{B} \sqrt{1 - \xi_\mathrm{B}^2} \sin (\delta_\mathrm{B}) \cos (\delta_\mathrm{B}) }{\sin^2 (\delta_\mathrm{B}) + \xi_\mathrm{B}^2 \cos^2 (\delta_\mathrm{B}) } \right).
\end{align}
Note that in the absence of noise, the ABP method can perfectly estimate the angle, i.e., $\widehat{\nu}_{\mathrm{d, t}} = \nu_{\mathrm{d, t}}$. Then, the target's horizontal angle from the perspective of the BS is
\begin{equation}
    \widehat{\psi}_\mathrm{d, t} = \arcsin \left( \frac{ \widehat{\nu}_\mathrm{d, t}}{\pi} \right).
\end{equation}
In this subsection, we assume that the BS deploys the ULA antenna structure; however, the ABP method can be similarly applied when the BS antennas and RIS elements are configured in the UPA structures as in our system model in Section~\ref{sec: System model}.

\subsection{SNR Improvement}
Before utilizing the ABP method to estimate the target angles from the perspectives of the BS and RIS, we first want to emphasize that the SNR of the received echo signals can be improved in our partial search procedure, which can lead to enhanced accuracy.
In the partial search, during each period where the UE receive codeword varies from $\bv^{(1)}$ to $\bv^{(L)}$, the BS transmit codeword and the RIS reflection codeword remain fixed, as illustrated in Fig.~\ref{Fig:Beam_training}.
Specifically, for each $k = 1, \dots, N_\mathrm{B} + N_\mathrm{R}$, during the $n$-th time slot where $n = (k-1)L+1, \dots, kL$, the same codeword combination is used as
\begin{align}
    &\widetilde{\bff}_{k} = \bff_{(k-1)L+1} = \cdots = \bff_{kL}, \nonumber \\ &\widetilde{\bw}_{k} = \bw_{(k-1)L+1} = \cdots = \bw_{kL}, \nonumber \\ &\widetilde{\boldsymbol{\Phi}}_{k} = \boldsymbol{\Phi}_{(k-1)L+1} = \cdots = \boldsymbol{\Phi}_{kL}.
\end{align}
During these time slots, the received echo signal at the BS in \eqref{echo signal} can be expressed as
\begin{align} \label{echo signal period}
    y_{\mathrm{s}, n} & = \sqrt{P_\mathrm{T}} \beta_\mathrm{t} \widetilde{\bw}_k^\mathrm{H} \big( \bh_{\mathrm{d, t}} + \bH_{\mathrm{BR}}^\mathrm{H} \widetilde{\boldsymbol{\Phi}}_k^\mathrm{H} \bh_{\mathrm{R, t}} \big) \nonumber\\  & \quad \times \big( \bh_{\mathrm{d, t}}^\mathrm{H} + \bh_{\mathrm{R, t}}^\mathrm{H} \widetilde{\boldsymbol{\Phi}}_k \bH_{\mathrm{BR}} \big) \widetilde{\bff}_k + \widetilde{\bw}_k^\mathrm{H} \bn_{\mathrm{s}, n} \nonumber \\& = \sqrt{P_\mathrm{T}} \beta_\mathrm{t} \widetilde{\bw}_k^\mathrm{H} \widetilde{\bh}_{\mathrm{eff}, k} \widetilde{\bh}_{\mathrm{eff}, k}^\mathrm{H} \widetilde{\bff}_k + \widetilde{\bw}_k^\mathrm{H} \bn_{\mathrm{s}, n},
\end{align}
where we define $\widetilde{\bh}_{\mathrm{eff}, k}^\mathrm{H} \triangleq \bh_{\mathrm{d, t}}^\mathrm{H} + \bh_{\mathrm{R, t}}^\mathrm{H} \widetilde{\boldsymbol{\Phi}}_k \bH_{\mathrm{BR}}$ as the effective channel from the BS to the target.
As in \eqref{echo signal period}, since the UE receive codeword $\bv_n$ does not affect the echo signal, the desired signal component remains constant throughout this period.
To leverage this property, we combine all the received echo signals over this period, which is defined as
\begin{align} \label{received echo sum}
    \widetilde{y}_{\mathrm{s}, k} &\triangleq \sum_{n = (k-1)L + 1}^{kL} y_{\mathrm{s}, n} \nonumber \\ &= \sum_{n = (k-1)L + 1}^{kL} \sqrt{P_\mathrm{T}} \beta_\mathrm{t} \widetilde{\bw}_k^\mathrm{H} \widetilde{\bh}_{\mathrm{eff}, k} \widetilde{\bh}_{\mathrm{eff}, k}^\mathrm{H} \widetilde{\bff}_k + \widetilde{\bw}_k^\mathrm{H} \bn_{\mathrm{s}, n} \nonumber \\ &= L \sqrt{P_\mathrm{T}} \beta_\mathrm{t} \widetilde{\bw}_k^\mathrm{H} \widetilde{\bh}_{\mathrm{eff}, k} \widetilde{\bh}_{\mathrm{eff}, k}^\mathrm{H} \widetilde{\bff}_k + \sum_{n = (k-1)L + 1}^{kL} \widetilde{\bw}_k^\mathrm{H} \bn_{\mathrm{s}, n}.
\end{align}
Note that the desired signal components are coherently combined, increasing the signal power by a factor of $L^2$.
In contrast, the noise power increases by a factor of $L$ because the independent noise terms are incoherently added.
The received SNR is improved by a factor of $L$, and therefore we use the coherently combined echo signals $\{ \widetilde{y}_{\mathrm{s}, k } \}_{k=1}^{N_\mathrm{B} + N_\mathrm{R}}$ for angle estimation in the following subsection.

\subsection{Angle Estimation from the Perspectives of the BS and RIS}\label{subsec: Angle Estimation}
To estimate the target's angles from the perspective of the BS, we adopt the ABP method for our system model in Section~\ref{sec: System model}.
In addition, with the properly designed RIS codebook in Section~\ref{subsec: Codebook construction}, we demonstrate that the ABP method can also be applied to the RIS reflection codewords to estimate the target's angles from the perspective of the RIS.
Each of these estimations is performed using the signals from the BS-target direct link and the BS-RIS-target reflection link.
However, since the BS receives both signals simultaneously as in \eqref{echo signal}, the reflection link can interfere with the accurate estimation of angles from the perspective of the BS and vice versa.
To mitigate this issue, we leverage the asymptotic orthogonality of array response vectors, which can ensure that one link becomes dominant in each estimation process.

During the first stage of the partial search, we estimate the target's angles from the perspective of the BS based on the coherently combined echo signals in \eqref{received echo sum}.
Since both vertical and horizontal angles will be estimated, we need to construct the vertical and horizontal ABPs.
First, we select the BS transmit codeword $\widetilde{\bff}_{k}$ that achieves the largest signal power among the $N_\mathrm{B}$ coherently combined echo signals $\{ \widetilde{y}_{\mathrm{s}, k } \}_{k=1}^{N_\mathrm{B}}$, which is denoted as $\ba_\mathrm{B} ( \gamma_\mathrm{B}^\mathrm{max}, \omega_\mathrm{B}^\mathrm{max} )$.
To form the vertical ABP, we compare the powers of the coherently combined echo signals when using two vertically adjacent beams, i.e., $\ba_\mathrm{B} ( \gamma_\mathrm{B}^\mathrm{max} - 2 \delta_\mathrm{B, v}, \omega_\mathrm{B}^\mathrm{max} )$ and $ \ba_\mathrm{B} ( \gamma_\mathrm{B}^\mathrm{max} + 2 \delta_\mathrm{B, v}, \omega_\mathrm{B}^\mathrm{max} )$, and identify the beam with the larger signal power.
Then, we define the boresight of the vertical ABP as $\overline{\gamma}_\mathrm{B}$, and each beam in the vertical ABP is represented as $\ba_\mathrm{B} ( \overline{\gamma}_\mathrm{B} - \delta_\mathrm{B, v}, \omega_\mathrm{B}^\mathrm{max} )$ and $\ba_\mathrm{B} ( \overline{\gamma}_\mathrm{B} + \delta_\mathrm{B, v}, \omega_\mathrm{B}^\mathrm{max} )$.
Since the beams in the vertical ABP are aligned with the target, not the RIS, when we use one of the vertical ABP beams $\widetilde{\bff}_{k} = \ba_\mathrm{B} ( \overline{\gamma}_\mathrm{B} - \delta_\mathrm{B, v}, \omega_\mathrm{B}^\mathrm{max} )$, the following approximation holds~\cite{Orthogonal_1, Orthogonal_2}
\begin{align} \label{approx}
    \bH_{\mathrm{BR}} \widetilde{\bff}_{k} &= \alpha_\mathrm{BR} \ba_\mathrm{R} ( \gamma_\mathrm{RB}, \omega_\mathrm{RB} ) \ba_\mathrm{B}^\mathrm{H} ( \gamma_\mathrm{BR}, \omega_\mathrm{BR} ) \nonumber \\ & \ \ \times \ba_\mathrm{B} ( \overline{\gamma}_\mathrm{B} - \delta_\mathrm{B, v}, \omega_\mathrm{B}^\mathrm{max} ) \underset{N \rightarrow \infty}{\approx} \boldsymbol{0}_{M}.
\end{align}
A similar approximation $\widetilde{\bw}_{k}^\mathrm{H} \bH_\mathrm{BR}^\mathrm{H} \underset{N \rightarrow \infty}{\approx} \boldsymbol{0}_{M}^\mathrm{T}$ also holds since we use the same BS transmit and receive codewords during the beam training.
Neglecting the noise and applying the asymptotic orthogonality as in \eqref{approx}, the coherently combined echo signal can be expressed as
\begin{align}
    \widetilde{y}_\mathrm{B, v}^\Delta &\approx L \sqrt{P_\mathrm{T}} \beta_\mathrm{t} \widetilde{\bw}_{k}^\mathrm{H} \Big( \bh_{\mathrm{d, t}} \bh_{\mathrm{d, t}}^\mathrm{H} +\bH_{\mathrm{BR}}^\mathrm{H} \widetilde{\boldsymbol{\Phi}}_k^\mathrm{H} \bh_{\mathrm{R, t}} \bh_{\mathrm{d, t}}^\mathrm{H} \nonumber \\ & \ \ + \bh_{\mathrm{d, t}} \bh_{\mathrm{R, t}}^\mathrm{H} \widetilde{\boldsymbol{\Phi}}_k \bH_{\mathrm{BR}} + \bH_{\mathrm{BR}}^\mathrm{H} \widetilde{\boldsymbol{\Phi}}_k^\mathrm{H} \bh_{\mathrm{R, t}} \bh_{\mathrm{R, t}}^\mathrm{H} \widetilde{\boldsymbol{\Phi}}_k \bH_{\mathrm{BR}} \Big) \widetilde{\bff}_{k} \nonumber \\ &\approx L \sqrt{P_\mathrm{T}} \beta_\mathrm{t} \widetilde{\bw}_{k}^\mathrm{H} \bh_{\mathrm{d, t}} \bh_{\mathrm{d, t}}^\mathrm{H} \widetilde{\bff}_{k},
\end{align}
which indicates that the signal from the BS-target link becomes dominant.
Then, the magnitude of $\widetilde{y}_\mathrm{B, v}^\Delta$ is denoted as
\begin{align}
    &\vert \widetilde{y}_\mathrm{B, v}^\Delta \vert \nonumber \\ &\approx \left\vert L \sqrt{P_\mathrm{T}} \beta_\mathrm{t} \widetilde{\bw}_{k}^\mathrm{H} \bh_{\mathrm{d, t}} \bh_{\mathrm{d, t}}^\mathrm{H} \widetilde{\bff}_{k} \right\vert \nonumber \\ &= \underbrace{L \sqrt{P_\mathrm{T}} \vert \beta_\mathrm{t} \vert \vert \alpha_\mathrm{d, t} \vert^2}_{\triangleq \beta_\mathrm{d, t}} \left\vert \ba_\mathrm{B}^\mathrm{H} ( \overline{\gamma}_\mathrm{B} - \delta_\mathrm{B, v}, \omega_\mathrm{B}^\mathrm{max} ) \ba_\mathrm{B} ( \gamma_{\mathrm{d, t}}, \omega_{\mathrm{d, t}} ) \right\vert^2 \nonumber \\ &= \beta_\mathrm{d, t} \left\vert \frac{1}{N_\mathrm{v}} \sum_{k=0}^{N_\mathrm{v}-1} e^{j k (\gamma_{\mathrm{d, t}} - \overline{\gamma}_\mathrm{B} + \delta_\mathrm{B, v}) }\right\vert^2 \nonumber \\ & \ \ \times \left\vert \frac{1}{N_\mathrm{h}} \sum_{m=0}^{N_\mathrm{h}-1} e^{j m (\omega_{\mathrm{d, t}} - \omega_\mathrm{B}^\mathrm{max}) }\right\vert^2 \nonumber \\ &= \beta_\mathrm{d, t} \frac{\cos^2 \left( \frac{N_\mathrm{v}(\gamma_{\mathrm{d, t}} - \overline{\gamma}_\mathrm{B})}{2} \right)}{N_\mathrm{v}^2 \sin^2 \left( \frac{\gamma_{\mathrm{d, t}} - \overline{\gamma}_\mathrm{B} + \delta_\mathrm{B, v}}{2} \right)} \frac{ \sin^2 \left( \frac{ N_\mathrm{h} ( \omega_{\mathrm{d, t}} - \omega_\mathrm{B}^\mathrm{max} ) }{2} \right)}{N_\mathrm{h}^2 \sin^2 \left( \frac{\omega_{\mathrm{d, t}} - \omega_\mathrm{B}^\mathrm{max} }{2} \right)}.
\end{align}
Similarly, when we use the other beam in the vertical ABP $\widetilde{\bff}_{k} = \ba_\mathrm{B} ( \overline{\gamma}_\mathrm{B} + \delta_\mathrm{B, v}, \omega_\mathrm{B}^\mathrm{max} )$, the coherently combined echo signal magnitude can be approximated as
\begin{equation}
    \vert \widetilde{y}_\mathrm{B, v}^\Sigma \vert \approx \beta_\mathrm{d, t} \frac{\cos^2 \left( \frac{N_\mathrm{v}(\gamma_{\mathrm{d, t}} - \overline{\gamma}_\mathrm{B})}{2} \right)}{N_\mathrm{v}^2 \sin^2 \left( \frac{\gamma_{\mathrm{d, t}} - \overline{\gamma}_\mathrm{B} - \delta_\mathrm{B, v}}{2} \right)} \frac{\sin^2 \left( \frac{ N_\mathrm{h} ( \omega_{\mathrm{d, t}} - \omega_\mathrm{B}^\mathrm{max} ) }{2} \right)}{N_\mathrm{h}^2 \sin^2 \left( \frac{\omega_{\mathrm{d, t}} - \omega_\mathrm{B}^\mathrm{max} }{2} \right)}.
\end{equation}
As in \eqref{ratio metric}, the ratio metric $\xi_\mathrm{B, v}$ can be defined as
\begin{equation} \label{ratio_metric_B_v}
    \xi_\mathrm{B, v} = \frac{\vert \widetilde{y}_\mathrm{B, v}^\Delta \vert - \vert \widetilde{y}_\mathrm{B, v}^\Sigma \vert}{\vert \widetilde{y}_\mathrm{B, v}^\Delta \vert + \vert \widetilde{y}_\mathrm{B, v}^\Sigma \vert} = - \frac{\sin(\gamma_\mathrm{d, t} - \overline{\gamma}_\mathrm{B}) \sin(\delta_\mathrm{B, v})}{1 - \cos(\gamma_\mathrm{d, t} - \overline{\gamma}_\mathrm{B}) \cos(\delta_\mathrm{B, v}) }.
\end{equation}
If the target lies within the range of vertical ABP, i.e., $\vert \gamma_\mathrm{d, t} - \overline{\gamma}_\mathrm{B} \vert \le \delta_\mathrm{B, v}$, the ratio metric $\xi_\mathrm{B, v}$ is a monotonically decreasing and invertible function of $\gamma_\mathrm{d, t} - \overline{\gamma}_\mathrm{B}$.
After taking the inverse function of \eqref{ratio_metric_B_v}, the vertical arrival spatial frequency of the BS-target link is estimated by
\begin{align} \label{gamma_d_t}
    \widehat{\gamma}_{\mathrm{d, t}} &= \overline{\gamma}_\mathrm{B} - \arcsin\Bigg( \frac{\xi_\mathrm{B, v} \sin (\delta_\mathrm{B, v})}{\sin^2 (\delta_\mathrm{B, v}) + \xi_\mathrm{B, v}^2 \cos^2 (\delta_\mathrm{B, v}) } \nonumber \\& \ \ - \frac{\xi_\mathrm{B, v} \sqrt{1 - \xi_\mathrm{B, v}^2} \sin (\delta_\mathrm{B, v}) \cos (\delta_\mathrm{B, v}) }{\sin^2 (\delta_\mathrm{B, v}) + \xi_\mathrm{B, v}^2 \cos^2 (\delta_\mathrm{B, v}) } \Bigg).
\end{align}
By the definition of vertical spatial frequency, the vertical angle from the perspective of the BS is estimated by
\begin{equation} \label{varphi_d_t}
    \widehat{\varphi}_\mathrm{d, t} = \arcsin \left( \frac{ \widehat{\gamma}_\mathrm{d, t}}{\pi} \right).
\end{equation}

Similar to the vertical angle estimation process, we construct the horizontal ABP.
Based on $\ba_\mathrm{B} ( \gamma_\mathrm{B}^\mathrm{max}, \omega_\mathrm{B}^\mathrm{max} )$, which maximizes the coherently combined echo signal in the first stage, we select the beam with the larger signal power among two horizontally adjacent beams.
Let the boresight of the horizontal ABP be defined as $\overline{\omega}_\mathrm{B}$.
Then, each beam in the horizontal ABP can be expressed as $\ba_\mathrm{B} ( \gamma_\mathrm{B}^\mathrm{max},  \overline{\omega}_\mathrm{B} - \delta_\mathrm{B, h} )$ and $\ba_\mathrm{B} ( \gamma_\mathrm{B}^\mathrm{max}, \overline{\omega}_\mathrm{B} + \delta_\mathrm{B, h} )$.
The coherently combined echo signals in \eqref{received echo sum} when using the horizontal ABP beams can be approximated as 
\begin{align} \label{horizontal ABP approx}
    \vert \widetilde{y}_\mathrm{B, h}^\Delta \vert &\approx  \beta_\mathrm{d, t} \frac{\sin^2 \left( \frac{N_\mathrm{v}(\gamma_{\mathrm{d, t}} - \gamma_\mathrm{B}^\mathrm{max})}{2} \right)}{N_\mathrm{v}^2 \sin^2 \left( \frac{\gamma_{\mathrm{d, t}} - \gamma_\mathrm{B}^\mathrm{max}}{2} \right)} \frac{\cos^2 \left( \frac{ N_\mathrm{h} ( \omega_{\mathrm{d, t}} - \overline{\omega}_\mathrm{B} ) }{2} \right)}{N_\mathrm{h}^2 \sin^2 \left( \frac{\omega_{\mathrm{d, t}} - \overline{\omega}_\mathrm{B} + \delta_\mathrm{B, h} }{2} \right)}, \nonumber \\ \vert \widetilde{y}_\mathrm{B, h}^\Sigma \vert &\approx \beta_\mathrm{d, t} \frac{\sin^2 \left( \frac{N_\mathrm{v}(\gamma_{\mathrm{d, t}} - \gamma_\mathrm{B}^\mathrm{max})}{2} \right)}{N_\mathrm{v}^2\sin^2 \left( \frac{\gamma_{\mathrm{d, t}} - \gamma_\mathrm{B}^\mathrm{max}}{2} \right)} \frac{\cos^2 \left( \frac{ N_\mathrm{h} ( \omega_{\mathrm{d, t}} - \overline{\omega}_\mathrm{B} ) }{2} \right)}{N_\mathrm{h}^2 \sin^2 \left( \frac{\omega_{\mathrm{d, t}} - \overline{\omega}_\mathrm{B} - \delta_\mathrm{B, h} }{2} \right)}.
\end{align}
With the approximation results in \eqref{horizontal ABP approx}, the ratio metric $\xi_\mathrm{B, h}$ is denoted by
\begin{equation} \label{ratio_metric_B_h}
    \xi_\mathrm{B, h} = \frac{\vert \widetilde{y}_\mathrm{B, h}^\Delta \vert - \vert \widetilde{y}_\mathrm{B, h}^\Sigma \vert}{\vert \widetilde{y}_\mathrm{B, h}^\Delta \vert + \vert \widetilde{y}_\mathrm{B, h}^\Sigma \vert} = - \frac{\sin(\omega_\mathrm{d, t} - \overline{\omega}_\mathrm{B}) \sin(\delta_\mathrm{B, h})}{1 - \cos(\omega_\mathrm{d, t} - \overline{\omega}_\mathrm{B}) \cos(\delta_\mathrm{B, h}) }.
\end{equation}
Assuming $\vert \omega_\mathrm{d, t} - \overline{\omega}_\mathrm{B} \vert \le \delta_\mathrm{B, h}$, the horizontal arrival spatial frequency of the BS-target link is estimated as
\begin{align} \label{omega_d_t}
    \widehat{\omega}_{\mathrm{d, t}} &= \overline{\omega}_\mathrm{B} - \arcsin\Bigg( \frac{\xi_\mathrm{B, h} \sin (\delta_\mathrm{B, h})}{\sin^2 (\delta_\mathrm{B, h}) + \xi_\mathrm{B, h}^2 \cos^2 (\delta_\mathrm{B, h}) } \nonumber \\ & \ \ - \frac{\xi_\mathrm{B, h} \sqrt{1 - \xi_\mathrm{B, h}^2} \sin (\delta_\mathrm{B, h}) \cos (\delta_\mathrm{B, h}) }{\sin^2 (\delta_\mathrm{B, h}) + \xi_\mathrm{B, h}^2 \cos^2 (\delta_\mathrm{B, h}) } \Bigg).
\end{align}
Then, the horizontal angle from the perspective of the BS can be estimated by using the vertical and horizontal spatial frequencies as
\begin{equation} \label{theta_d_t}
    \widehat{\theta}_\mathrm{d, t} = \arcsin \left( \frac{\widehat{\omega}_\mathrm{d, t}}{\pi} \left( 1 - \left( \frac{ \widehat{\gamma}_\mathrm{d, t}}{\pi } \right)^2 \right)^{-\frac{1}{2}} \right).
\end{equation}

In the second stage of the partial search, we estimate the target's vertical and horizontal angles from the perspective of the RIS by applying the ABP method to the RIS reflection codewords.
As in the above explanation, the first thing to do is to find the RIS reflection codeword that maximizes the coherently combined echo signals in $\{ \widetilde{y}_{\mathrm{s}, k } \}_{k=N_\mathrm{B}+1}^{N_\mathrm{B}+N_\mathrm{R}}$, and it is denoted by $\widetilde{\boldsymbol{\phi}}_k = \sqrt{M} \ba_\mathrm{R} ( \gamma_\mathrm{R}^\mathrm{max} - \gamma_\mathrm{RB}, \omega_\mathrm{R}^\mathrm{max} - \omega_\mathrm{RB})$.
To form the vertical ABP, we compare the powers of the coherently combined echo signals when using two vertically adjacent RIS reflection codewords, i.e., $\sqrt{M} \ba_\mathrm{R} ( \gamma_\mathrm{R}^\mathrm{max} - \gamma_\mathrm{RB} - 2 \delta_\mathrm{R, v}, \omega_\mathrm{R}^\mathrm{max} - \omega_\mathrm{RB} )$ and $ \sqrt{M} \ba_\mathrm{R} ( \gamma_\mathrm{R}^\mathrm{max} - \gamma_\mathrm{RB} + 2 \delta_\mathrm{R, v}, \omega_\mathrm{R}^\mathrm{max} - \omega_\mathrm{RB} )$.
After selecting the reflection codeword with the larger signal power, we denote the boresight of the vertical ABP as $\overline{\gamma}_\mathrm{R}$.
Then, the RIS reflection codewords in the vertical ABP are represented as $\sqrt{M} \ba_\mathrm{R} ( \overline{\gamma}_\mathrm{R} - \gamma_\mathrm{RB} - \delta_\mathrm{R, v}, \omega_\mathrm{R}^\mathrm{max} - \omega_\mathrm{RB} )$ and $\sqrt{M} \ba_\mathrm{R} ( \overline{\gamma}_\mathrm{R} - \gamma_\mathrm{RB} + \delta_\mathrm{R, v}, \omega_\mathrm{R}^\mathrm{max} - \omega_\mathrm{RB} )$.
Note that the terms $\gamma_\mathrm{RB}$ and $\omega_\mathrm{RB}$ remain in the expression because they cancel out when compensating for the phase of the BS-RIS link.
During the second stage, since the BS transmit codeword is steered toward the direction of the RIS, the following approximation holds~\cite{Orthogonal_1, Orthogonal_2}
\begin{equation}
    \bh_{\mathrm{d, t}}^\mathrm{H} \widetilde{\bff}_k =  \alpha_{\mathrm{d, t}}^\mathrm{H} \ba_\mathrm{B}^\mathrm{H} ( \gamma_{\mathrm{d, t}}, \omega_{\mathrm{d, t}} ) \ba_\mathrm{B} ( \gamma_\mathrm{BR}, \omega_\mathrm{BR} ) \underset{N \rightarrow \infty}{\approx} 0,
\end{equation}
and similarly, the approximation $\widetilde{\bw}_k^\mathrm{H} \bh_{\mathrm{d, t}} \underset{N \rightarrow \infty}{\approx} 0$ also holds. By exploiting this asymptotic orthogonality and neglecting the noise, the coherently combined echo signal in the second stage can be approximated as
\begin{align}
    \widetilde{y}_{\mathrm{s}, k} &\approx L \sqrt{P_\mathrm{T}} \beta_\mathrm{t} \widetilde{\bw}_k^\mathrm{H} \Big( \bh_{\mathrm{d, t}} \bh_{\mathrm{d, t}}^\mathrm{H} +\bH_{\mathrm{BR}}^\mathrm{H} \widetilde{\boldsymbol{\Phi}}_k^\mathrm{H} \bh_{\mathrm{R, t}} \bh_{\mathrm{d, t}}^\mathrm{H} \nonumber \\ & \ \ + \bh_{\mathrm{d, t}} \bh_{\mathrm{R, t}}^\mathrm{H} \widetilde{\boldsymbol{\Phi}}_k \bH_{\mathrm{BR}} + \bH_{\mathrm{BR}}^\mathrm{H} \widetilde{\boldsymbol{\Phi}}_k^\mathrm{H} \bh_{\mathrm{R, t}} \bh_{\mathrm{R, t}}^\mathrm{H} \widetilde{\boldsymbol{\Phi}}_k \bH_{\mathrm{BR}}  \Big) \widetilde{\bff}_k \nonumber \\ &\approx L \sqrt{P_\mathrm{T}} \beta_\mathrm{t} \widetilde{\bw}_k^\mathrm{H} \bH_{\mathrm{BR}}^\mathrm{H} \widetilde{\boldsymbol{\Phi}}_k^\mathrm{H} \bh_{\mathrm{R, t}} \bh_{\mathrm{R, t}}^\mathrm{H} \widetilde{\boldsymbol{\Phi}}_k \bH_{\mathrm{BR}} \widetilde{\bff}_k,
\end{align}
which shows that the BS-RIS-target reflection link is dominant in this stage.

The magnitude of the coherently combined echo signal is then given by
\begin{align}
     \vert \widetilde{y}_{\mathrm{s}, k} \vert &\approx \left\vert L \sqrt{P_\mathrm{T}} \beta_\mathrm{t} \widetilde{\bw}_k^\mathrm{H} \bH_{\mathrm{BR}}^\mathrm{H} \widetilde{\boldsymbol{\Phi}}_k^\mathrm{H} \bh_{\mathrm{R, t}} \bh_{\mathrm{R, t}}^\mathrm{H} \widetilde{\boldsymbol{\Phi}}_k \bH_{\mathrm{BR}} \widetilde{\bff}_k \right\vert \nonumber \\ & = \underbrace{L \sqrt{P_\mathrm{T}} \vert \beta_\mathrm{t} \vert \vert \alpha_\mathrm{BR} \vert^2 \vert \alpha_\mathrm{R, t} \vert^2}_{\triangleq \beta_\mathrm{R, t}} \Big\vert \ba_\mathrm{B}^\mathrm{H} ( \gamma_\mathrm{BR}, \omega_\mathrm{BR} ) \ba_\mathrm{B}( \gamma_\mathrm{BR}, \omega_\mathrm{BR} ) \nonumber \\& \ \ \times \ba_\mathrm{R}^\mathrm{H} ( \gamma_\mathrm{RB}, \omega_\mathrm{RB} ) \widetilde{\boldsymbol{\Phi}}_k^\mathrm{H}  \ba_\mathrm{R} ( \gamma_{\mathrm{R, t}}, \omega_{\mathrm{R, t}} ) \Big\vert^2 \nonumber \\ & = \beta_\mathrm{R, t} \left\vert \ba_\mathrm{R}^\mathrm{H} ( \gamma_\mathrm{RB}, \omega_\mathrm{RB} ) \widetilde{\boldsymbol{\Phi}}_k^\mathrm{H}  \ba_\mathrm{R} ( \gamma_{\mathrm{R, t}}, \omega_{\mathrm{R, t}} ) \right\vert^2 \nonumber \\ & \stackrel{\mathrm{(a)}}{=} \beta_\mathrm{R, t} \left\vert \widetilde{\boldsymbol{\phi}}_k^\mathrm{H} \diag \left( \ba_\mathrm{R}^\mathrm{H} ( \gamma_\mathrm{RB}, \omega_\mathrm{RB} ) \right)  \ba_\mathrm{R} ( \gamma_{\mathrm{R, t}}, \omega_{\mathrm{R, t}} ) \right\vert^2,
\end{align}
where (a) follows from the definition of $\widetilde{\boldsymbol{\Phi}}_k = \diag(\widetilde{\boldsymbol{\phi}}_k)$. Note that $\beta_\mathrm{R, t}$ includes the channel gains of both the BS-RIS and RIS-target channels and thus reflects the multiplicative path-loss effect in RIS-aided systems.\footnote{Due to this effect, the target angle estimation performance from the perspective of the RIS can be worse than that from the perspective of the~BS.}
When the RIS reflection codeword are set to $\widetilde{\boldsymbol{\phi}}_k = \sqrt{M} \ba_\mathrm{R} ( \overline{\gamma}_\mathrm{R} - \gamma_\mathrm{RB} - \delta_\mathrm{R, v}, \omega_\mathrm{R}^\mathrm{max} - \omega_\mathrm{RB} )$, which is in the vertical ABP, the magnitude of the coherently combined echo signal is given by
\begin{align} \label{echo signal approx a}
    &\vert \widetilde{y}_\mathrm{R, v}^\Delta \vert \nonumber \\ &\approx \beta_\mathrm{R, t} \big\vert \sqrt{M} \ba_\mathrm{R}^\mathrm{H} ( \overline{\gamma}_\mathrm{R} - \gamma_\mathrm{RB} -  \delta_\mathrm{R, v}, \omega_\mathrm{R}^\mathrm{max} - \omega_\mathrm{RB} ) \nonumber \\ & \ \ \times \diag ( \ba_\mathrm{R}^\mathrm{H} ( \gamma_\mathrm{RB}, \omega_\mathrm{RB} ) ) \ba_\mathrm{R} ( \gamma_{\mathrm{R, t}}, \omega_{\mathrm{R, t}} ) \big\vert^2 \nonumber \\ &= \beta_\mathrm{R, t} \left\vert \ba_\mathrm{R}^\mathrm{H} ( \overline{\gamma}_\mathrm{R} - \delta_\mathrm{R, v}, \omega_\mathrm{R}^\mathrm{max} ) \ba_\mathrm{R} ( \gamma_{\mathrm{R, t}}, \omega_{\mathrm{R, t}} ) \right\vert^2 \nonumber \\ &=  \beta_\mathrm{R, t} \left\vert \frac{1}{M_\mathrm{v}} \sum_{k=0}^{M_\mathrm{v}-1} e^{j k (\gamma_{\mathrm{R, t}} - \overline{\gamma}_\mathrm{R} + \delta_\mathrm{R, v}) }\right\vert^2 \nonumber \\ & \ \ \times \left\vert \frac{1}{M_\mathrm{h}}\sum_{m=0}^{M_\mathrm{h}-1} e^{j m (\omega_{\mathrm{R, t}} - \omega_\mathrm{R}^\mathrm{max}) }\right\vert^2 \nonumber \\ &= \beta_\mathrm{R, t} \frac{\cos^2 \left( \frac{M_\mathrm{v}(\gamma_{\mathrm{R, t}} - \overline{\gamma}_\mathrm{R})}{2} \right)}{M_\mathrm{v}^2\sin^2 \left( \frac{\gamma_{\mathrm{R, t}} - \overline{\gamma}_\mathrm{R} + \delta_\mathrm{R, v}}{2} \right)} \frac{\sin^2 \left( \frac{ M_\mathrm{h} ( \omega_{\mathrm{R, t}} - \omega_\mathrm{R}^\mathrm{max} ) }{2} \right)}{M_\mathrm{h}^2 \sin^2 \left( \frac{\omega_{\mathrm{R, t}} - \omega_\mathrm{R}^\mathrm{max} }{2} \right)}.
\end{align}
With the other reflection codeword in the vertical ABP $\widetilde{\boldsymbol{\phi}}_k = \sqrt{M} \ba_\mathrm{R} ( \overline{\gamma}_\mathrm{R} - \gamma_\mathrm{RB} + \delta_\mathrm{R, v}, \omega_\mathrm{R}^\mathrm{max} - \omega_\mathrm{RB} )$, the coherently combined echo signal magnitude is similarly expressed as
\begin{align} \label{echo signal approx b}
    \vert \widetilde{y}_\mathrm{R, v}^\Sigma \vert &\approx \beta_\mathrm{R, t} \frac{\cos^2 \left( \frac{M_\mathrm{v}(\gamma_{\mathrm{R, t}} - \overline{\gamma}_\mathrm{R})}{2} \right)}{M_\mathrm{v}^2 \sin^2 \left( \frac{\gamma_{\mathrm{R, t}} - \overline{\gamma}_\mathrm{R} - \delta_\mathrm{R, v}}{2} \right)} \frac{\sin^2 \left( \frac{ M_\mathrm{h} ( \omega_{\mathrm{R, t}} - \omega_\mathrm{R}^\mathrm{max} ) }{2} \right)}{M_\mathrm{h}^2 \sin^2 \left( \frac{\omega_{\mathrm{R, t}} - \omega_\mathrm{R}^\mathrm{max} }{2} \right)}.
\end{align}
By applying \eqref{echo signal approx a} and \eqref{echo signal approx b}, we define the ratio metric $\xi_\mathrm{R, v}$ as in \eqref{ratio_metric_B_v}.
Then, the vertical arrival spatial frequency of the RIS-target link and the vertical angle from the perspective of the RIS can be estimated in the same manner as in \eqref{gamma_d_t} and \eqref{varphi_d_t}, and they are denoted by $\widehat{\gamma}_{\mathrm{R, t}}$ and $\widehat{\varphi}_\mathrm{R, t}$, respectively.

To estimate the horizontal angle from the perspective of the RIS, we construct the horizontal ABP of the RIS reflection codewords.
Among the two candidates horizontally adjacent to $\sqrt{M} \ba_\mathrm{R} ( \gamma_\mathrm{R}^\mathrm{max} - \gamma_\mathrm{RB}, \omega_\mathrm{R}^\mathrm{max} - \omega_\mathrm{RB})$, we select the one with the larger power of the coherently combined echo signal.
The RIS reflection codewords in the horizontal ABP can be expressed as $\sqrt{M} \ba_\mathrm{R} ( \gamma_\mathrm{R}^\mathrm{max} - \gamma_\mathrm{RB}, \overline{\omega}_\mathrm{R} - \omega_\mathrm{RB} - \delta_\mathrm{R, h} )$ and $\sqrt{M} \ba_\mathrm{R} ( \gamma_\mathrm{R}^\mathrm{max} - \gamma_\mathrm{RB}, \overline{\omega}_\mathrm{R} - \omega_\mathrm{RB} +  \delta_\mathrm{R, h} )$ where $\overline{\omega}_\mathrm{R}$ denote the boresight of the horizontal ABP.
The coherently combined echo signals when using the RIS reflection codewords in the horizontal ABP are denoted as $\widetilde{y}_\mathrm{R, h}^\Delta$ and $\widetilde{y}_\mathrm{R, h}^\Sigma$.
As in \eqref{ratio_metric_B_h}, the ratio metric $\xi_\mathrm{R, h}$ can be similarly defined. The horizontal arrival spatial frequency of the RIS-target link and the horizontal angle from the perspective of the RIS can be estimated as in \eqref{omega_d_t} and \eqref{theta_d_t}. These estimates are denoted by $\widehat{\omega}_{\mathrm{R, t}}$ and $\widehat{\theta}_\mathrm{R, t}$, respectively.

\subsection{Target Localization} \label{subsec: Target localization}

\begin{figure}[t]
    \centering
    \includegraphics[width=0.7 \columnwidth]{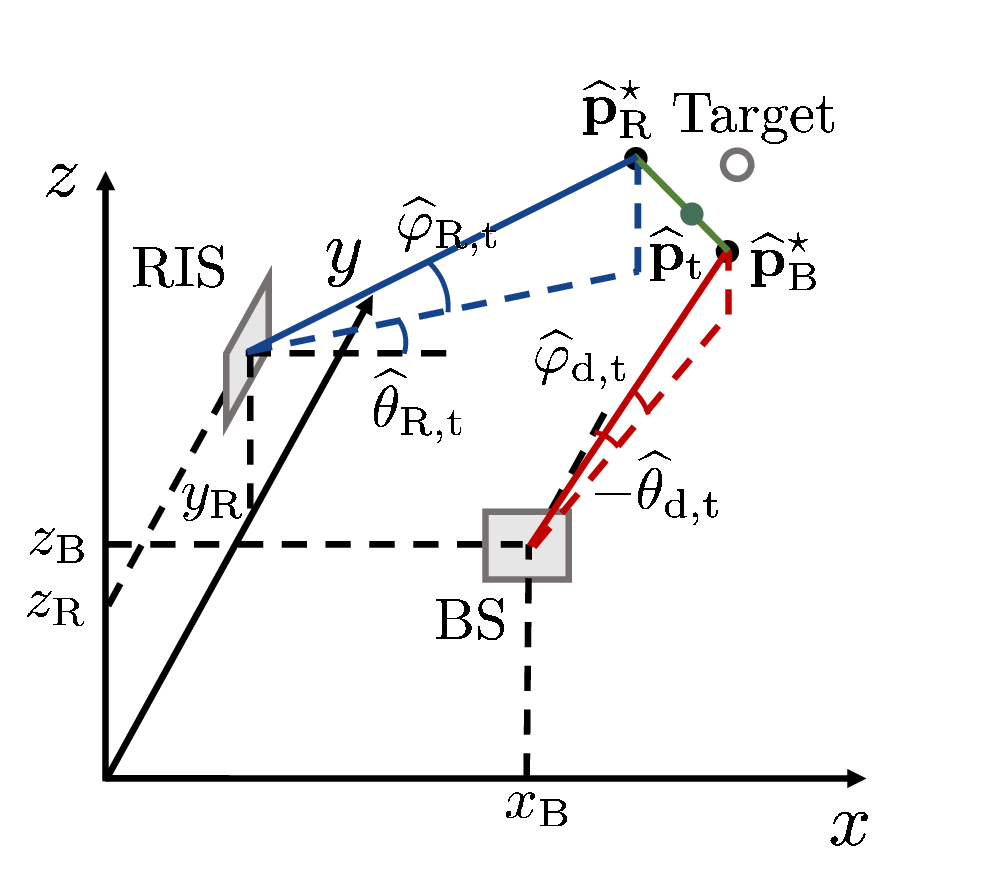}
    \caption{Geometric representation of the proposed target localization technique.} 
    \label{Fig:Localization_UPA}
\end{figure}

In this subsection, based on the estimated angles, we propose a closed-form target localization technique.
Considering the 3D coordinate system as shown in Fig.~\ref{Fig:Localization_UPA}, we place the BS along the x-axis and the RIS along the y-axis.
The BS antennas and RIS elements are aligned parallel to the xz-plane and yz-plane.
The position vectors of the BS and RIS are given by $\bl_\mathrm{B} = ( x_\mathrm{B}, 0, z_\mathrm{B} )^\mathrm{T}$ and $\bl_\mathrm{R} = ( 0, y_\mathrm{R}, z_\mathrm{R} )^\mathrm{T}$.
Based on geometric relationships, the normalized direction vectors from the BS and RIS to the target are expressed as
\begin{align}
    &\widehat{\bd}_\mathrm{B} = \left( - \cos (\widehat{\varphi}_\mathrm{d, t} ) \sin (\widehat{\theta}_\mathrm{d, t} ), \cos (\widehat{\varphi}_\mathrm{d, t} ) \cos (\widehat{\theta}_\mathrm{d, t} ), \sin (\widehat{\varphi}_\mathrm{d, t} ) \right)^\mathrm{T}, \nonumber \\ & \widehat{\bd}_\mathrm{R} = \left( \cos (\widehat{\varphi}_\mathrm{R, t} ) \cos ( \widehat{\theta}_\mathrm{R, t} ), \cos (\widehat{\varphi}_\mathrm{R, t} ) \sin ( \widehat{\theta}_\mathrm{R, t} ), \sin (\widehat{\varphi}_\mathrm{R, t} ) \right)^\mathrm{T}.
\end{align}
Then, the lines extending from the BS and RIS in the directions of estimated angles are defined as
\begin{equation}
    \widehat{\bp}_\mathrm{B} = \bl_\mathrm{B} + t_\mathrm{B} \widehat{\bd}_\mathrm{B}, \ \widehat{\bp}_\mathrm{R} = \bl_\mathrm{R} + t_\mathrm{R} \widehat{\bd}_\mathrm{R},
\end{equation}
where $t_\mathrm{B} \in \mathbb{R}^+$ and $t_\mathrm{R} \in \mathbb{R}^+$ are parameters, which represent the distances along the corresponding direction vectors.
If the angles are perfectly estimated, the two lines will intersect at the target's position; however, due to the inevitable angle estimation errors, the lines do not intersect in general.
Therefore, we aim to find the closest pair of points lying on the two lines.

Let us denote the closest points on the directional lines from the BS and RIS as $\widehat{\bp}_\mathrm{B}^\star$ and $\widehat{\bp}_\mathrm{R}^\star$, respectively, which correspond to the parameters $t_\mathrm{B}^\star$ and $t_\mathrm{R}^\star$.
To compute these parameters, we utilize the property that the vector connecting the closest points, i.e., $\widehat{\bp}_\mathrm{B}^\star - \widehat{\bp}_\mathrm{R}^\star$,  is orthogonal to both direction vectors, which can be expressed as follows
\begin{align} \label{equation}
    &( \widehat{\bp}_\mathrm{B}^\star - \widehat{\bp}_\mathrm{R}^\star ) ^\mathrm{T} \widehat{\bd}_\mathrm{B} = ( \bl_\mathrm{B} - \bl_\mathrm{R} + t_\mathrm{B}^\star \widehat{\bd}_\mathrm{B} - t_\mathrm{R}^\star \widehat{\bd}_\mathrm{R} )^\mathrm{T} \widehat{\bd}_\mathrm{B} = 0, \nonumber \\ &( \widehat{\bp}_\mathrm{B}^\star - \widehat{\bp}_\mathrm{R}^\star )^\mathrm{T} \widehat{\bd}_\mathrm{R} = ( \bl_\mathrm{B} - \bl_\mathrm{R} + t_\mathrm{B}^\star \widehat{\bd}_\mathrm{B} - t_\mathrm{R}^\star \widehat{\bd}_\mathrm{R} )^\mathrm{T} \widehat{\bd}_\mathrm{R} = 0.  
\end{align}
By solving the linear equations in \eqref{equation}, the parameters $t_\mathrm{B}^\star$ and $t_\mathrm{R}^\star$ can be computed as
\begin{align}
    &t_\mathrm{B}^\star = \frac{((\widehat{\bd}_\mathrm{B}^\mathrm{T} \widehat{\bd}_\mathrm{R}) \widehat{\bd}_\mathrm{R} - \widehat{\bd}_\mathrm{B})^\mathrm{T} (\bl_\mathrm{B} - \bl_\mathrm{R})}{1 - (\widehat{\bd}_\mathrm{B}^\mathrm{T} \widehat{\bd}_\mathrm{R})^2 }, \nonumber \\ & t_\mathrm{R}^\star = \frac{(\widehat{\bd}_\mathrm{R} - ( \widehat{\bd}_\mathrm{B} ^\mathrm{T} \widehat{\bd}_\mathrm{R} ) \widehat{\bd}_\mathrm{B} )^\mathrm{T} (\bl_\mathrm{B} - \bl_\mathrm{R})}{1 - (\widehat{\bd}_\mathrm{B}^\mathrm{T} \widehat{\bd}_\mathrm{R})^2 }.
\end{align}
After finding the closest points, we estimate the target position by using an internal division point between them with appropriate weights, which is given by
\begin{equation}
    \widehat{\bp}_\mathrm{t} = \frac{\lambda_1}{\lambda_1 + \lambda_2} \widehat{\bp}_\mathrm{B}^\star + \frac{\lambda_2}{\lambda_1 + \lambda_2} \widehat{\bp}_\mathrm{R}^\star, \label{eq: estimated position}
\end{equation}
where $\lambda_1$ and $\lambda_2$ denote the weight parameters that can be properly designed.
If the angle estimation from the perspective of the BS is more accurate than that from the perspective of the RIS, $\widehat{\bd}_\mathrm{B}$ becomes more reliable, and hence $\widehat{\bp}_\mathrm{B}^\star$ should be assigned a larger weight and vice versa.
Since the maximum magnitudes of the coherently combined echo signals obtained in the first and second stages are directly related to the SNR of the signals used for angle estimation from the perspectives of the BS and RIS, respectively, we set $\lambda_1$ and $\lambda_2$~as 
\begin{equation} \label{weight}
    \lambda_1 = \max_{k = 1, \cdots, N_\mathrm{B}} \vert \widetilde{y}_{\mathrm{s}, k} \vert, \ \lambda_2 = \max_{k = N_\mathrm{B} + 1, \cdots, N_\mathrm{B} + N_\mathrm{R}} \vert \widetilde{y}_{\mathrm{s}, k} \vert.
\end{equation}

\subsection{Extension to Multi-Target Scenario} \label{subsec: Multi-target}
To demonstrate that the proposed technique is extensible to multi-target localization, here we consider the case of two point-like targets for ease of explanation. Note that the extension to an arbitrary number of targets is straightforward. Then, the coherently combined echo signals in \eqref{received echo sum} become
\begin{align}
    \widetilde{y}_{\mathrm{s}, k} &\triangleq L \sqrt{P_\mathrm{T}} \sum_{g=1}^2 \beta_{\mathrm{t}, g} \widetilde{\bw}_k^\mathrm{H} \widetilde{\bh}_{\mathrm{eff}, k, g} \widetilde{\bh}_{\mathrm{eff}, k, g}^\mathrm{H} \widetilde{\bff}_k \nonumber \\ & \quad + \sum_{n = (k-1)L + 1}^{kL} \widetilde{\bw}_k^\mathrm{H} \bn_{\mathrm{s}, n},
\end{align}
where $\widetilde{\bh}_{\mathrm{eff}, k, g}^\mathrm{H} \triangleq \bh_{\mathrm{d, t}, g}^\mathrm{H} + \bh_{\mathrm{R, t},g}^\mathrm{H} \widetilde{\boldsymbol{\Phi}}_k \bH_{\mathrm{BR}}$ is the effective channel from the BS to the $g$-th target, for $g \in \{1,2\}$. The channels from the BS and the RIS to the $g$-th target are denoted by $\bh_{\mathrm{d, t}, g}^\mathrm{H}$ and $\bh_{\mathrm{R, t}, g}^\mathrm{H}$, respectively, and $\beta_{\mathrm{t}, g} \sim \cC \cN(0, 1)$ is the normalized RCS of the $g$-th target.

During the first stage, we apply the ABP method following the procedure in Section~\ref{subsec: Angle Estimation} to estimate the target angles from the perspective of the BS. These angle estimates correspond to the target with the dominant channel; for clarity, we assume that this is the first target. Then, we reconstruct the signal reflected from the first target and remove its impact from the coherently combined echo signals during the first stage. Let $k^\star$ be the index that maximizes the power of the coherently combined echo signals, i.e., $k^\star = \argmax_{k=1,\dots,N_\mathrm{B}} \vert \widetilde{y}_{\mathrm{s}, k } \vert$. The coherently combined echo signal at index $k = k^\star$ can be approximated as
\begin{align}
    \widetilde{y}_{\mathrm{s}, k^\star} \nonumber &\approx L \sqrt{P_\mathrm{T}} \beta_{\mathrm{t}, 1} \widetilde{\bw}_{k^\star}^\mathrm{H} \bh_{\mathrm{d, t}, 1} \bh_{\mathrm{d, t}, 1}^\mathrm{H} \widetilde{\bff}_{k^\star} \nonumber \\ &= L \sqrt{P_\mathrm{T}} \underbrace{\beta_{\mathrm{t}, 1} \vert \alpha_{\mathrm{d, t}, 1} \vert^2}_{\triangleq \ \eta_{\mathrm{d, t}, 1}} \left\vert \ba_\mathrm{B}^\mathrm{H} (\gamma_{\mathrm{d, t}, 1}, \omega_{\mathrm{d, t}, 1}) \widetilde{\bff}_{k^\star} \right\vert^2,
\end{align}
with $\bh_{\mathrm{d, t}, 1} = \alpha_{\mathrm{d, t}, 1} \ba_\mathrm{B} (\gamma_{\mathrm{d, t}, 1}, \omega_{\mathrm{d, t}, 1})$ where $\gamma_{\mathrm{d, t}, 1}$ and $\omega_{\mathrm{d, t}, 1}$ are the vertical and horizontal arrival spatial frequencies.
Since we have already obtained $\widehat{\gamma}_{\mathrm{d, t}, 1}$ and $\widehat{\omega}_{\mathrm{d, t}, 1}$ by applying the ABP method, we can estimate $\eta_{\mathrm{d, t}, 1}$, which captures the target RCS and the complex-valued channel gain, as follows
\begin{align}
    \widehat{\eta}_{\mathrm{d, t}, 1} &= \arg\min_\eta \left\vert \widetilde{y}_{\mathrm{s}, k^\star} - \eta L \sqrt{P_\mathrm{T}} \vert \ba_\mathrm{B}^\mathrm{H} (\widehat{\gamma}_{\mathrm{d, t}, 1}, \widehat{\omega}_{\mathrm{d, t}, 1}) \widetilde{\bff}_{k^\star} \vert^2 \right\vert_2^2 \nonumber \\ &= \frac{\widetilde{y}_{\mathrm{s}, k^\star}}{L \sqrt{P_\mathrm{T}} \vert \ba_\mathrm{B}^\mathrm{H} (\widehat{\gamma}_{\mathrm{d, t}, 1}, \widehat{\omega}_{\mathrm{d, t}, 1}) \widetilde{\bff}_{k^\star} \vert^2}.
\end{align}
Then, for all coherently combined echo signals in the first stage, we can remove the first target's impact as
\begin{equation}
    \check{y}_{\mathrm{s}, k} = \widetilde{y}_{\mathrm{s}, k} - L \sqrt{P_\mathrm{T}} \widehat{\eta}_{\mathrm{d, t}, 1} \vert \ba_\mathrm{B}^\mathrm{H} (\widehat{\gamma}_{\mathrm{d, t}, 1}, \widehat{\omega}_{\mathrm{d, t}, 1}) \widetilde{\bff}_{k} \vert^2.
\end{equation}
Using the updated echo signals $\{ \check{y}_{\mathrm{s}, k } \}_{k=1}^{N_\mathrm{B}}$, the second target's angles from the perspective of the BS can be estimated by applying the ABP method, with $\widetilde{y}_{\mathrm{s}, k }$ replaced by $\check{y}_{\mathrm{s}, k }$. As a result, we can obtain two sets of angles from the perspective of the BS. Similarly, during the second stage, we adopt the ABP method to estimate the angles of the dominant target from the perspective of the RIS and remove its impact from the coherently combined echo signals. Then, we reapply the ABP method to estimate the angles of the other target.

In practical scenarios, it is difficult to determine which target each estimated angle corresponds to, so the estimated angle sets from the perspective of the BS and those from the perspective of the RIS must be properly matched. From the BS and RIS, two lines are obtained for each viewpoint, extending along the corresponding estimated angle directions. We search for the pair with the shortest distance between the two lines, one originating from the BS and the other from the RIS, and identify this pairing as one target. The remaining pair is assigned to the other target. After matching the angle sets between the BS and the RIS, the target localization can be performed as in Section~\ref{subsec: Target localization}.

\section{Numerical Results} \label{sec: Numerical results}

\begin{figure*}[h]
     \centering
     \begin{subfigure}[t]{0.32\linewidth}
         \centering
         \includegraphics[width=\textwidth]{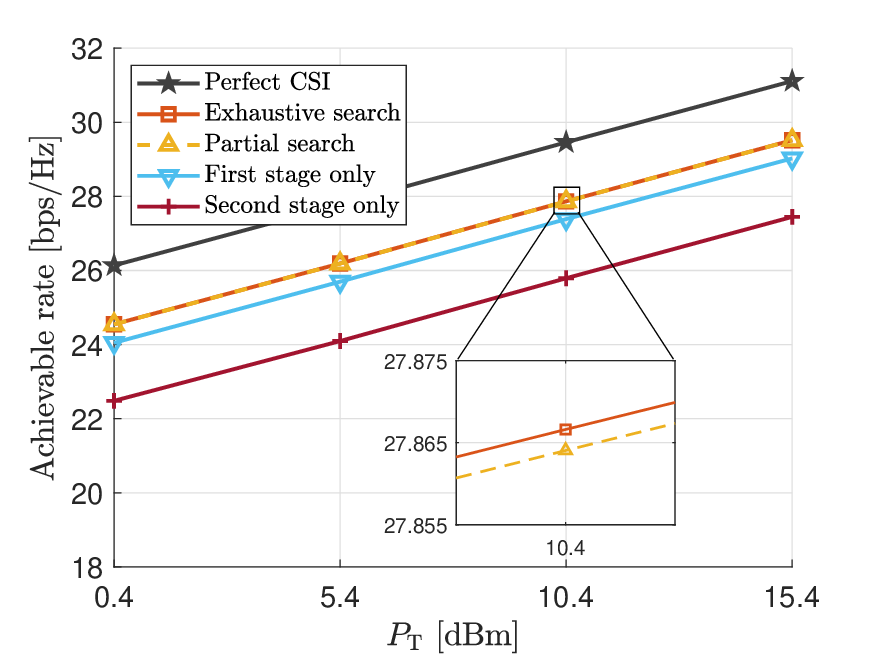}
         \caption{}
         \label{Fig:Rate_UPA_BS_Power}
     \end{subfigure}
     \hfill
     \begin{subfigure}[t]{0.32\linewidth}
         \centering
         \includegraphics[width=\textwidth]{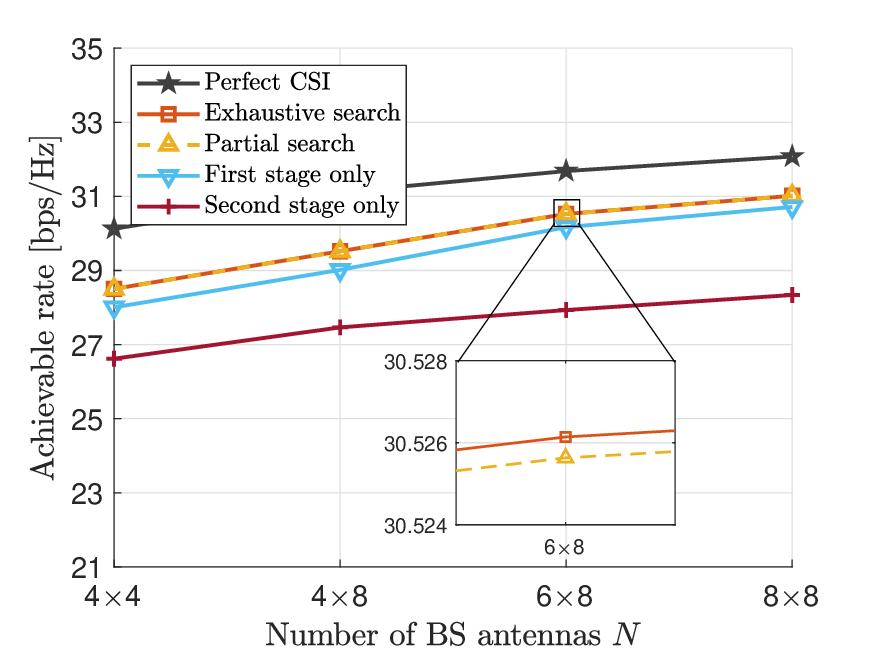}
         \caption{}
         \label{Fig:Rate_UPA_BS_antennas}
     \end{subfigure}
     \hfill
     \begin{subfigure}[t]{0.32\linewidth}
         \centering
         \includegraphics[width=\textwidth]{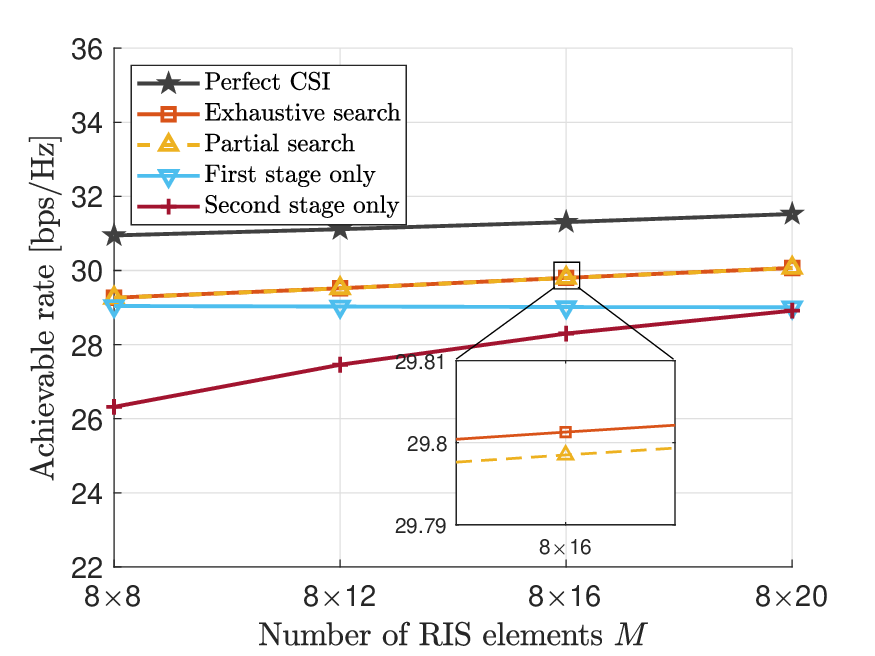}
         \caption{}
         \label{Fig:Rate_UPA_RIS_elements}
     \end{subfigure}
        \caption{Achievable rate performances according to the different values of $P_\mathrm{T}$, $N$, and $M$.}
        \label{fig: UPA-Rate}
\end{figure*}

In this section, we evaluate the performance of the proposed technique described in Sections \ref{sec: Beam training process} and \ref{sec: Proposed} in terms of both sensing and communication metrics for the ISAC systems. The numerical environments are constructed as follows. For a 3D coordinate system as in Fig. \ref{Fig:Localization_UPA}, we assume the fixed locations for the BS and RIS where they are located at $(10 \ \text{m}, 0 \ \text{m}, 15 \ \text{m})$ and $(0 \ \text{m}, 20 \ \text{m}, 10 \ \text{m})$. However, the locations of the single point-like target and the communication UE are randomly generated. When we denote location of target as $(x_\mathrm{t} \ \text{m}, y_\mathrm{t} \ \text{m}, z_\mathrm{t} \ \text{m} )$, it is assumed that $x_\mathrm{t} \sim \mathrm{Unif}(6, 10), y_\mathrm{t} \sim \mathrm{Unif}(20, 24)$, and $ z_\mathrm{t} \sim \mathrm{Unif}(11, 15)$ where $\mathrm{Unif}(a,b)$ is the uniform distribution for the interval $[a,b]$. Similarly, the location of UE is denoted as $(x_\mathrm{c} \ \mathrm{m}, y_\mathrm{c} \ \mathrm{m}, 1 \ \mathrm{m})$ where $x_\mathrm{c} \sim \mathrm{Unif}(10, 20)$ and $ y_\mathrm{c} \sim \mathrm{Unif}(20, 30)$. 

To consider a narrowband scenario, we assume that a single subcarrier is used in the OFDM system. Under this setup, we adopt the Rician fading model with one LoS path and multiple NLoS paths for all channel links as in \cite{Rician_1, Rician_2}. The path-loss at the reference distance $1 \ \text{m}$ is set to $-30 \ \text{dB}$, and the path-loss exponents for the BS-target (UE), BS-RIS, and RIS-target (UE) links are assumed to be 2.8, 2.1, and 2.2, respectively. The Rician K-factor and the number of NLoS paths are the same for all links as 7 dB and 4. The arrival and departure angles of LoS path are numerically obtained with the actual locations, and the angles of NLoS paths are randomly generated with vertical angular spread $5^\circ$ and horizontal angular spread $8^\circ$ centered at the LoS path. Considering the total bandwidth $10 \ \text{MHz}$, the subcarrier spacing is set to $30 \ \text{kHz}$. The number of resource blocks is $24$ according to the 5G NR specification~\cite{3GPP_TS_38101_1}, where each resource block consists of $12$ subcarriers. Then, the total number of active subcarriers is $288$. With the noise power spectral density $-174 \ \text{dBm/Hz}$, the noise variances at the UE and BS over one subcarrier are set as $\sigma_\mathrm{c}^2 = \sigma_\mathrm{s}^2 = - 129.2 \ \text{dBm}$. Unless otherwise stated, we assume $N = 4\times 8$ BS antennas, $M = 8\times 12$ RIS elements, and $L = 4$ UE antennas. The total transmit power is set to $40 \ \text{dBm}$, and by equally allocating the transmit power over all active subcarriers, the transmit power for each subcarrier is set to $P_\mathrm{T} = 15.4 \ \text{dBm}$.

\subsection{Communication Rate Performance}
Following the beam training procedure and codebook design in Section \ref{sec: Beam training process}, the most appropriate codeword combination will be selected to maximize the received signal power at the UE. Although the codebook design is based on the well-known approach that utilizes the structure of the 5G standard, it is still necessary to evaluate the communication performance at the UE in terms of a proper communication metric, e.g., the achievable rate. Through the downlink received signal model in \eqref{eq: received signal at the UE}, the achievable rate $R$ with the BS beamforming vector $\bff$, the UE beamforming vector $\bv$, and the reflection coefficient matrix $\boldsymbol{\Phi}$ at the RIS is given by
\begin{equation}
    R=\log_2 \left( 1 + \frac{ P_\mathrm{T} \vert \bv^\mathrm{H} \left( \bH_{\mathrm{d, c}} + \bH_{\mathrm{R, c}} \boldsymbol{\Phi} \bH_{\mathrm{BR}}\right) \bff \vert^2  }{\sigma_\mathrm{c}^2} \right). \label{eq: achievable rate}
\end{equation}

As described in Section \ref{sec: Beam training process}, we considered two different training procedures: exhaustive search and partial search. To find the optimal codeword combination of $\{\bff_n, \boldsymbol{\Phi}_n, \bv_n\}$ that maximizes the power of the received signal at the UE in~\eqref{eq: received signal at the UE}, the exhaustive search procedure will consider in total $N_\mathrm{B} N_\mathrm{R} L$ possible candidates, and only the $(N_\mathrm{B} + N_\mathrm{R})L$ candidates will be explored for the partial search procedure. In addition, we consider two baseline beam training procedures that perform only the first stage or the second stage of the partial search, exploring $N_\mathrm{B} L$ and $N_\mathrm{R} L$ candidates, respectively. Then, the achievable rate for each case is determined by applying the optimal codeword combination to \eqref{eq: achievable rate}. Since the obtained optimal codeword combinations do not directly maximize the achievable rate, we adopt a baseline scheme in \cite{Baseline_RIS_perfect_CSI} that serves as an upper bound by directly maximizing the achievable rate of an RIS-aided MIMO communication system. Assuming perfect channel state information (CSI), the method in \cite{Baseline_RIS_perfect_CSI} employs an alternating optimization framework to obtain effective non-codebook-based beamforming vectors at the BS and UE, as well as the reflection coefficients at the RIS. 

Fig. \ref{fig: UPA-Rate} shows the average achievable rate performances of the baseline perfect CSI case and the other beam training procedures according to the different values of parameters $P_\mathrm{T}$, $N$, and $M$. 
Even with lower training overhead, the partial search procedure can have a similar achievable rate to the exhaustive search procedure. This implies that the performance degradation due to considering only a portion of beam candidates is negligible regardless of the parameter values, as we discussed in Section~\ref{sec: Beam training process}. A reasonable achievable rate can be obtained by performing only the first or second stage; however, for these cases, the target angle can be estimated only from the perspective of the BS or the RIS, which makes target localization impossible. The achievable rate difference compared to the perfect CSI case is small when using the codeword combination from the designed codebooks, indicating that the proposed codebooks properly cover the spatial domain. Although the baseline perfect CSI case outperforms the others, it is not clear how the target localization can be performed for this ideal case. Because codebooks are designed to enable not only the beam alignment for the communication UE but also the high-resolution target localization, the results in Fig. \ref{fig: UPA-Rate} emphasize the versatility of the codebook design for the beam training procedure.

\subsection{Target Localization Performance}
To evaluate the sensing performance, we adopt the normalized mean squared error (NMSE) of the target position relative to the BS as the performance metric, which is given by
\begin{align}
    \mathrm{NMSE} &= \mathbb{E}\left[\frac{\Vert (\bp_{\mathrm{t}} - \bl_\mathrm{B}) - (\widehat{\bp}_{\mathrm{t}} - \bl_\mathrm{B}) \Vert_2^2}{\Vert \bp_{\mathrm{t}} - \bl_\mathrm{B} \Vert_2^2}\right] =  \mathbb{E}\left[\frac{\Vert \bp_{\mathrm{t}} - \widehat{\bp}_{\mathrm{t}} \Vert_2^2}{\Vert \bp_{\mathrm{t}} - \bl_\mathrm{B} \Vert_2^2} \right] , 
\end{align}
where $\widehat{\bp}_{\mathrm{t}} - \bl_\mathrm{B}$ is the estimated target position relative to the BS. Based on this metric, we compare the proposed angle estimation and localization technique described in Section \ref{sec: Proposed} with the following benchmarks. 
\begin{itemize} 
    \item \textbf{Exhaustive search}: Choose the codeword combination whose power of the coherently combined echo signal is maximized among all possible candidates, and the estimated angles are obtained by the corresponding spatial frequencies of the chosen codewords at the BS and RIS. Based on the estimated angles, the proposed localization technique in Section \ref{subsec: Target localization} is applied.
    \item \textbf{Partial search}: The target angles from the perspective of the BS are obtained through the spatial frequencies of the codeword at the BS that maximizes the coherently combined echo signal power during the first stage. The angles from the perspective of the RIS are similarly obtained through the codeword at the RIS during the second stage.
    The localization is applied in the same way.
    \item \textbf{MUSIC+Grid}\footnote{Although the proposed design incurs slightly higher training overhead than the MUSIC+Grid case, it significantly reduces the computational complexity.}: The developed method in \cite{Baseline_MUSIC_Grid} considers a two-stage localization. The angle estimates from the perspective of the BS are obtained by applying a multiple signal classification (MUSIC)-based algorithm to the coherently combined echo signals in the first stage. Then, the on-grid angle estimation from the perspective of the RIS is applied. With the estimated angles, the localization is followed using the geometric relationship among the target, BS, and RIS.
    \item \textbf{MUSIC+Grid-proposed}: This case basically follows the angle estimation strategies in \cite{Baseline_MUSIC_Grid}, but the proposed localization technique based on the angle estimates is applied. By comparing with this case, the performance of the proposed localization technique itself can be explored. 
\end{itemize}

\begin{figure}[t]
    \centering
    \includegraphics[width=0.81 \columnwidth]{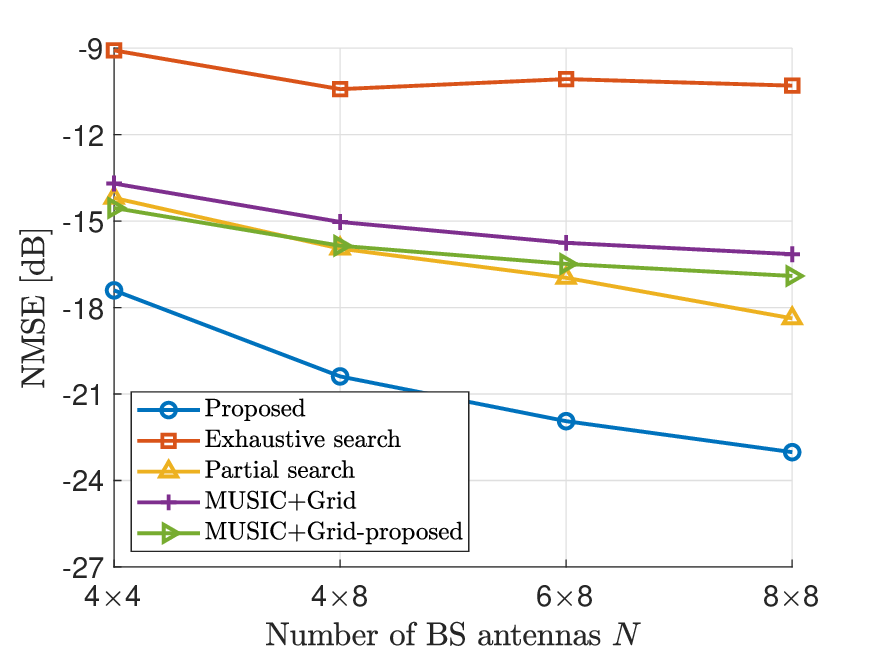}
    \caption{NMSE performance according to $N$.} 
    \label{Fig:UPA_NMSE_BS_antennas}
\end{figure}


Fig. \ref{Fig:UPA_NMSE_BS_antennas} shows the NMSE performance of the proposed technique and benchmarks according to the number of BS antennas $N$ as in Fig. \ref{Fig:Rate_UPA_BS_antennas}. It can be observed that the proposed design outperforms the benchmarks in terms of the NMSE, and the performance gap becomes much larger as $N$ increases. This is because the BS can generate narrower beams and exploit much higher-resolution angle estimation using the ABP method without incurring additional training overhead. The performance gap between the MUSIC+Grid case and the MUSIC+Grid-proposed case highlights the effectiveness of the proposed localization technique with the given angle estimates. It is worth noting that the localization method in \cite{Baseline_MUSIC_Grid} relies on geometric relationships, which can lead to position mismatches when the angle estimation is not perfect and the direction vectors from the perspectives of the BS and RIS cannot have an intersection. However, the proposed localization technique accounts for this possibility, and this robustness leads to better localization performance. The exhaustive search case shows worse NMSE performance than the partial search case, which may be counterintuitive. This is because when the coherently combined echo signal through the RIS is strong enough, the BS codeword is often steered toward the RIS to maximize the overall echo signal power rather than toward the target. This misalignment reduces the ability of the exhaustive search to correctly estimate the target angles, resulting in poor NMSE performance. Given that the proposed technique is based on the partial search procedure and that the achievable rate performance degradation in this case is negligible, it can be concluded that the proposed technique is effective from an ISAC perspective.

\begin{figure}[t]
    \centering
    \includegraphics[width=0.81 \columnwidth]{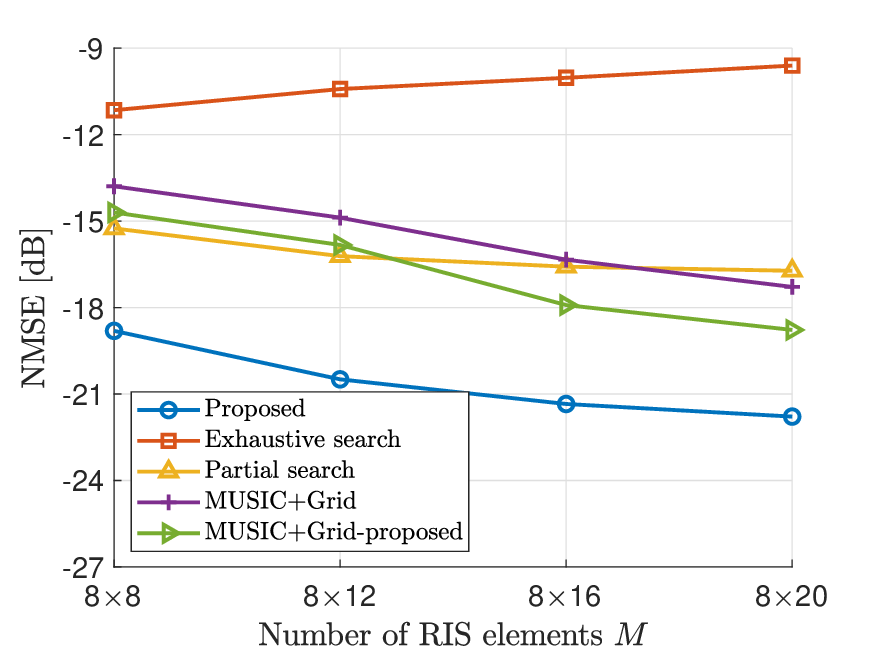}
    \caption{NMSE performance according to $M$.} 
    \label{Fig:UPA_NMSE_RIS_elements}
\end{figure}

Similar results can be shown through Fig. \ref{Fig:UPA_NMSE_RIS_elements} where the NMSE performances are compared according to the number of RIS elements $M$ as in Fig. \ref{Fig:Rate_UPA_RIS_elements}. The proposed technique can have the lowest NMSE performance regardless of the value of $M$, highlighting the advantages of the proposed design. While the performance improvement is less notable compared to the case of increasing the number of $N$, it is still effective because increasing the number of passive RIS elements is much more energy efficient. 

\begin{figure}[t]
    \centering
    \includegraphics[width=0.81 \columnwidth]{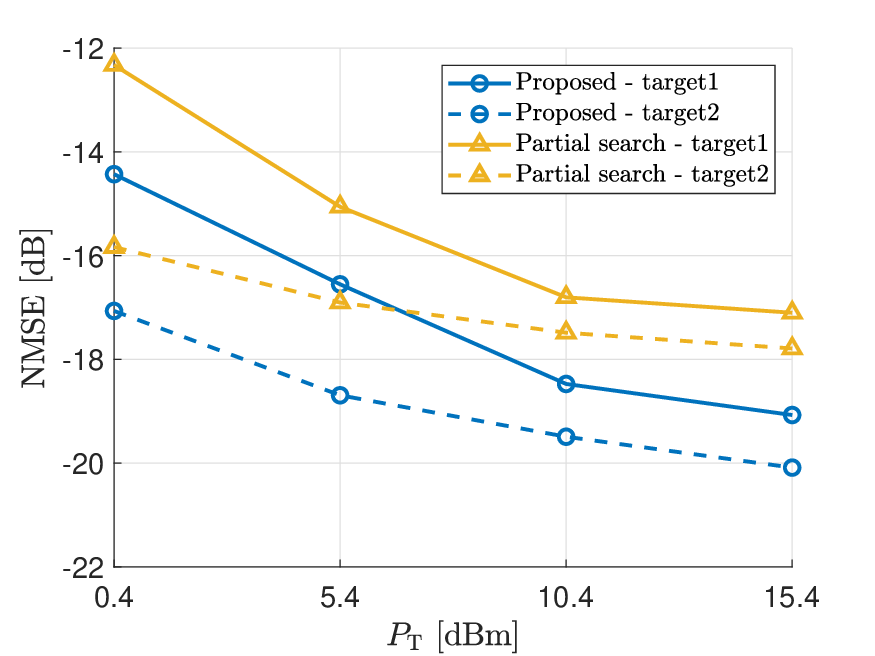}
    \caption{NMSE performance according to $P_\mathrm{T}$ in multi-target scenarios.} 
    \label{Fig:multi_target}
\end{figure}

In Fig. \ref{Fig:multi_target}, the NMSE performances according to the BS transmit power $P_\mathrm{T}$ in multi-target scenarios are depicted with $N = 8 \times 8$ BS antennas. The location of the first target is randomly generated as $x_\mathrm{t} \sim \mathrm{Unif}(3, 6), y_\mathrm{t} \sim \mathrm{Unif}(21, 24)$, and $ z_\mathrm{t} \sim \mathrm{Unif}(12, 15)$, whereas $x_\mathrm{t} \sim \mathrm{Unif}(3, 6), y_\mathrm{t} \sim \mathrm{Unif}(16, 19)$, and $ z_\mathrm{t} \sim \mathrm{Unif}(5, 8)$ are used for the second target. The NMSE performances of each target are shown separately. It can be observed that the multi-target localization extension explained in Section~\ref{subsec: Multi-target} works well, and for both targets, the proposed design outperforms the partial search case, which indicates that the ABP method is still effective in multi-target scenarios. In addition, as $P_\mathrm{T}$ increases, the performance gap between the proposed design and the partial search case increases because the ABP method further enhances the angular resolution as the impact of noise diminishes.

\section{Conclusion} \label{sec: Conclusion}
We developed the beam training framework for RIS-aided ISAC systems using codebooks designed according to the 5G standard. To reduce the training overhead, we proposed the two-stage partial search procedure. For the target sensing during the beam training procedure, we adopted the ABP method to estimate target angles from the perspectives of the BS and RIS with high resolution. Then, based on these estimates, we introduced a closed-form localization technique and further showed that the proposed framework can be extended to multi-target localization scenarios. Numerical results showed that the partial search procedure achieves communication rates comparable to the exhaustive search procedure and the perfect CSI case. In addition, it is also shown that the proposed target localization technique, which includes the angle estimation, outperforms the benchmarks. This demonstrates the effectiveness of the developed framework from an ISAC perspective.

Possible future research directions can include extending the scenario to multiple RIS deployments. Developing beam tracking algorithms applicable to UEs or targets with high mobility is another important future research topic. It is also worth investigating an extension of the proposed beam training framework that accounts for near-field channel characteristics when using a large number of BS antennas or RIS elements. Moreover, the multiplicative path-loss effect can degrade the angle estimation accuracy from the perspective of the RIS. To mitigate this limitation, deploying an active RIS is another interesting direction for future research. Although our framework enables target localization in narrowband scenarios, target delay remains important sensing information in ISAC systems. Therefore, extending the proposed framework to wideband systems by jointly exploiting angle and delay information will be also a promising research direction.


\bibliographystyle{IEEEtran}
\bibliography{refs_all}

\end{document}